\documentclass[a4paper]{aa}  

\usepackage{txfonts}
\usepackage[dvipsnames]{xcolor}
\usepackage{caption}
\usepackage{graphicx}
\usepackage{amsmath}
\usepackage{hyperref}
\usepackage[utf8]{inputenc}

\title{Vorticity and magnetic dynamo from subsonic expansion waves}

\authorrunning{Elias-López et al.}
\titlerunning{Vorticity and magnetic dynamo from subsonic expansion waves}
   \author{Albert Elias-López
          \inst{1,2},
          Fabio Del Sordo \inst{1,2,3}
          and
          Daniele Viganò \inst{1,2,4}
          }

\institute { Institute of Space Sciences (ICE-CSIC), Campus UAB, Carrer de Can Magrans s/n, 08193, Barcelona, Spain
\and
Institut d’Estudis Espacials de Catalunya (IEEC), 08034 Barcelona, Spain
\and
INAF, Osservatorio Astrofisico di Catania, via Santa Sofia, 78 Catania, Italy
\and
Institute of Applied Computing \& Community Code (IAC3), University of the Balearic Islands, Palma, 07122, Spain
\\
\email{albert.elias@csic.es}\\
\email{delsordo@ice.csic.es}\\
\email{daniele.vigano@csic.es}
             }

   \date{Received ---------; accepted --------}

  \abstract
   {The main driving forces supplying energy to the interstellar medium (ISM) are supernova explosions and stellar winds. Such localized sources are assimilable to curl-free velocity fields as a first approximation. They need to be combined with other physical processes to replicate real galactic environments, such as the presence of turbulence and a dynamo-sustained magnetic field in the ISM.}
   {This work is focused on the effect of an irrotational forcing on a magnetized flow in the presence of rotation, baroclinicity, shear, or a combination of any of the three. It follows an earlier analysis with a similar focus, namely, subsonic spherical expansion waves in hydrodynamic simulations. By including magnetic field in the model, we can evaluate the occurrence of dynamo on both small and large scales. We aim to identify the minimum ingredients needed to trigger a dynamo instability as well as the relation between dynamo and the growth of vorticity.}
   {We used the Pencil code to run resistive magnetohydrodynamic direct numerical simulations, exploring the ranges of values of several physical and numerical parameters of interest. We explored Reynolds numbers up to a few hundreds. We analyzed the temporal evolution of vorticity, kinetic, and magnetic energy, as well as their features in Fourier space.}
   {We report the absence of a small-scale dynamo in all cases where only rotation is included, regardless of the given equation of state and rotation rate. Conversely, the inclusion of a background sinusoidal shearing profile leads to an hydrodynamic instability that produces an exponential growth of the vorticity at all scales, starting from small ones. This is know as vorticity dynamo. The onset of this instability occurs after a rather long temporal evolution of  several thousand turbulent turnover times. The vorticity dynamo in turn drives an exponential growth of the magnetic field, first at small scales, followed by large ones. The instability is then saturated and the magnetic field approximately reaches equipartition with the turbulent kinetic energy. During the saturation phase, we can observe a winding of the magnetic field in the direction of the shearing flow. By varying the intensity of the shear, we see that the growth rates of this instability change. The inclusion of the baroclinic term has the main effect of delaying the onset of the vorticity dynamo, but then leads to a more rapid growth.}
   {Our work demonstrates how even purely irrotational forcing may lead to dynamo action in the presence of shear, thus amplifying the field to an equipartition level. At the same time, we confirm that purely irrotational forcing alone does not lead to any growth in terms of the vorticity, nor the magnetic field. This picture does not change in the presence of rotation or baroclinicity, at least up to a resolution of $256^3$ mesh points. To further generalize such a conclusion, we will need to explore how this setup works both at higher magnetic Reynolds numbers and with different prescriptions of the irrotational forcing.}


\begin{document} 

\maketitle
   
\section{Introduction} \label{Introduction}

Sources of energy input in the interstellar medium (ISM) are of stellar or galactic origin, and they lead the observed turbulence in the ISM. Supernova explosions (SNe), stellar winds, protostellar outflows, and even ionizing radiation (by decreasing order of importance) are known to contribute to injecting energy into the ISM. It is known that SNe are the main contributors of energy and can solely provide the energy needed to produce the turbulence observed in the ISM \citep[e.g.,][]{Gressel2008A&A...486L..35G}. They are also considered one of the main ingredients in determining galactic dynamos \citep[see][for a recent review]{BN2023ARAA}. Still, their effect on the amplification of galactic magnetic fields remains poorly understood. It is unclear how they contribute to the seeding and amplification of such fields to the observed values both for Milky Way-like galaxies \citep[e.g.,][]{Ntormousi2018A&A...619L...5N, Ntormousi2020A&A...641A.165N} and for high-redshift ones \citep{Ntormousi2022A&A...668L...6N}.  If they are seen as purely spherical, SNe should not (in principle) produce any vorticity in the surrounding medium.

However, there are many other aspects of the Galactic medium that ought to be taken into account. Stratification, differential rotation, shear, or shocks are examples of physical mechanisms which might help generate vortical flows from the available energy. Three-dimensional (3D) simulations of such environments can be quite demanding, especially considering the supersonic feature of the flow. This field was pioneered by \cite{Korpi1999ApJ...514L..99K}, who studied how SNe produce turbulence in a multiphase ISM, and then by \cite{Mac2005ApJ...626..864M}, who analyzed the distribution of pressure in a magnetized ISM. However, in these simulations, the generation of vorticity, which in turn is essential for amplifying the magnetic field, remains somewhat unclear. The question of whether vorticity can be amplified by a purely curl-free (compressible, or irrotational) forcing \citep[e.g.,][]{2010A&A...512A..81F, Federrathetal2011} or not \citep{MeeBrandenburg2006, DelSordoBrandenburg2011} is still
a matter of debate.

\cite{MeeBrandenburg2006} and \cite{DelSordoBrandenburg2011} used localized random expansion waves as forcing terms only in the momentum equation. This potential forcing induces spherically symmetric successive accelerations in random places during a given time interval, reaching locally trans-sonic regimes and leading to a somewhat similar SNe driving. Such an irrotationally forced flow was found not to generate any vorticity up to $512^3$ mesh points \citep{MeeBrandenburg2006}. However, when rotation, explicit shear or baroclinicity were added, \cite{DelSordoBrandenburg2011} found that vorticity is produced. Magnetic fields were added to the most simple setup (no rotation or shear, nor baroclinicity) by \cite{MeeBrandenburg2006}, who found no evidence of magnetic field amplification. On the contrary, they found that this type of flow has a highly destructive effect on the magnetic field at the smallest scales. In general, the role of SN feedback on regulating star formation rate and the structure of the galactic disk is still not completely understood \cite[e.g.,][]{2014A&A...570A..81H} and it may depend on how SNe are implemented in simulations. The nature of the forcing may have an effect on several features of the ISM, besides the growth of the magnetic field. For instance, \cite{2023MNRAS.518.5190M} showed how a purely compressible forcing in magnetohydrodynamic (MHD) simulations of star cluster formation influences the initial mass function. 

This work is a continuation of the research done by \cite{MeeBrandenburg2006} and \cite{DelSordoBrandenburg2011}.
We aim to study how shear, rotation, and baroclinicity, in combination with irrotational forcing in the form of localized random expansion waves, contribute to the decay or growth of an initial magnetic field seed. We want to explore the possible dynamo action in a parameter space defined by the forcing parameters, the rotation and shear rates, the Reynolds and Prandtl numbers, and the initial magnetic seed. Our work concentrates on the subsonic and transonic regime, with the aim of establishing  the conditions under which a purely compressible forcing may amplify vorticity and magnetic fields.
 
Other works have also studied the appearance of small-scale dynamo (SSD), which is dynamo action at length scales equal or smaller than the forcing scale, as a consequence of irrotational forcing. \cite{Porter2015ApJ...810...93P} performed  simulations of subsonic turbulence in intra-galactic cluster medium with a purely curl-free (compressible) forcing, finding that dynamo can be excited when divergence-free (solenoidal) modes arise. \cite{AchikanathFederrath2021} found turbulent dynamo in a highly subsonic regime by using a turbulent driving force in Fourier space either with solenoidal or with purely compressive contributions, or with a combination of them, following the approach of \cite{2010A&A...512A..81F} and \cite{Federrathetal2011}.

Other simulations have achieved magnetic field growth with more complex forcing of the same type. \cite{GentShukurov2013, GentShukurov2013Mag} simulated a multi-phase ISM randomly heated and stirred by SNe, in the energy equation, with a stratified, rotating and shearing local domain with a galactic gravitational potential and shock diffusivities. The obtained flow dynamics are robust when three major phases are used (defined with temperature ranges) and when the parameters are adjusted to reproduce individual SN remnants. \cite{KayplaGent2018} simulated such events with a combination of mass transfer during SNe and stellar formation, energy deposition in energy equation, and a stably stratified, rotating medium mimicking the galactic plane. They also obtained a SSD in such simulations. This was also confirmed in subsequent studies by \cite{2021ApJ...910L..15G, 2023ApJ...943..176G} of a multi-phase ISM. Recently, \cite{Seta2022MNRAS.514..957S} studied turbulent dynamo in a multi-phase ISM as well, providing a detailed analysis of vorticity sources in their simulations.

These models are fairly complex as they want to reproduce what happens in galactic environment. Our approach has instead been to work with a much simpler model to shed light on the minimum ingredients needed to amplify a magnetic field up to equipartition values in presence of a purely irrotational forcing. This is in the spirit of studying general aspects of dynamo generated magnetic fields and discuss their applications both in galactic environment, as well as in planetary and stellar ones. The analytical mean field approach by \cite{KrauseRadler1980mfmd.book.....K} and \cite{RS2006PhRvE} predicts the presence of an electromotive force and, hence, an amplified mean magnetic field, in the presence of a mean flow. While this is surely happening in the case of helical flows \cite[see e.g.,][]{BS2005PhR...417....1B, Rincon2019JPlPh..85d2001R}, a mean electromotive force can  (in principle) be observed also with non-helical turbulence in the presence of rotation \cite{Radler1968ZNatA..23.1851R,Radler1969WisBB..11..194R} or with a large-scale flow with associated vorticity \cite{Igor2003PhRvE..68c6301R}. Large-scale dynamos was found in numerical experiments employing a non-helical forcing to drive small-scale turbulence embedded in a large-scale shear flow \citep[e.g.,][]{Yousef2008AN....329..737Y,Brandenburg2008ApJ...676..740B, Singh2015ApJ...806..118S}. However, the forcing functions in these calculations did allow the injection of vorticity on small scales. In this work, we want instead to test the possibility of driving a dynamo process with turbulence driven in a purely compressible way, either in the absence or in the presence presence of large-scale flows.

This paper is organized as follows: in \S\ref{Numerical methods} we describe the MHD model and the organization of our numerical experiments. In \S\ref{Results} we show the results and in \S\ref{Astrophysical applications} we discuss the implication of these results in the framework of galactic dynamo. Finally, we present our conclusions in \S\ref{sec:conclusions}.

\section{Model and numerical methods} 
\label{Numerical methods}

\subsection{MHD equations with rotation and shear}
\label{sec:mhdeq}

\par
We ran numerical simulations using the Pencil Code, \cite{ThePencilCode}\footnote{https://github.com/pencil-code}. 
This is a non-conservative, high-order, finite-difference code (sixth-order accurate in space and third-order Runge-Kutta in time), which we employed to solve the non-ideal compressible MHD equations.

We considered either a rigidly rotating frame, with angular velocity $\mathbf{\Omega}=(0,0,\Omega)$ in the $z$-direction or a differential velocity (or shear) in the $y$-direction given by $\mathbf{u}^S = (0, u_y^S(z), 0)$. We did not consider stratification. We solved the equations in the co-rotating reference frame, for which the mass and momentum conservation equations are expressed as:
\begin{eqnarray}
 &&  \dfrac{D \ln \rho}{D t} = - \nabla \cdot \mathbf{u}~, \label{Continuity_equation}\\
&& \dfrac{D \mathbf{u}'}{D t} = -\frac{\nabla p}{\rho} + \frac{\mathbf{J} \times \mathbf{B}}{\rho} - 2 \mathbf{\Omega} \times \mathbf{u} - u_z\frac{\partial u_y^S}{\partial z}{\mathbf{\hat{y}}}  + \mathbf{F}_{\nu} + \mathbf{f} + \mathbf{f}_s,
\label{Equation_of_motion}
\end{eqnarray}
where: $\mathbf{u}(t)=(\mathbf{u}^S + \mathbf{u}'(t))$ is the total velocity field combining the fixed shearing velocity with the turbulent velocity $\mathbf{u}'$; $\rho$ is the mass density; $p$ is the pressure; $\mathbf{B}$ the magnetic field; and $\mathbf{J}$ the electrical current density $\mathbf{J} := (\nabla \times \mathbf{B})/\mu_0$ (where $\mu_0$ is the vacuum permeability).

Importantly, the advective derivative operator, $D/Dt := \partial / \partial t + \mathbf{u} \cdot \nabla$, applies to the total velocity, that is, to both the evolving turbulent velocity $\mathbf{u}'$ (kept in the left-hand side) and the fixed shearing velocity, $\mathbf{u}^S$. We keep the latter in the right-hand side, in the form of the only term $ - \mathbf{u}\cdot\nabla \mathbf{u^s} = - u_z \frac{\partial u^S_y}{\partial z}$. The other source terms are the viscous force, $\mathbf{F}_{\nu} = \rho^{-1} \nabla \cdot (2\rho \nu \mathbf{S})$, where the traceless rate of strain tensor $\mathbf{S}$ is $S_{ij} = (1/2)(u_{i,j}+u_{j,i} - (1/3) \delta_{ij} \nabla \cdot \mathbf{u})$, the external expansion wave forcing, $\mathbf{f,}$ and a shearing forcing, $\mathbf{f}_s$ (see below for their definitions).

In order to close the system of equations, we consider two types of equation of state (EoS): 1) a simple barotropic EoS $p(\rho)=c_s^2 \rho$, where we fix the value of the sound speed $c_s = 1$ and 2) an ideal EoS (dubbed also baroclinic case), $p(\rho,T)=\rho R_g T$, with $R_g$ the specific gas constant and $T$ the temperature; in this case, the sound speed squared is $c_s^2 = (\gamma - 1)c_p T$, where we fix the adiabatic index $\gamma = c_p/c_v = 5/3$ (corresponding to a monatomic perfect gas), and $c_p$ and $c_v$ are the specific heats at constant pressure and constant volume, respectively.

In the latter case, the energy equation is expressed in terms of $\rho$, the specific entropy, $s$, and $T$:

\begin{equation}
\label{Entropy_equation_Pencil}
\begin{aligned}
  T \dfrac{D s}{D t} = 2 \nu \boldsymbol{S} \otimes \boldsymbol{S} + \rho^{-1} \nabla (c_p \rho \chi \nabla T) + \rho^{-1} \eta \mu_0 \boldsymbol{J}^2 - \frac{1}{\tau_{cool}}(c_s^2 - c_{s0}^2)~,
\end{aligned}
\end{equation}
where $\chi$ is the thermal diffusivity, $\eta$ is the magnetic diffusivity, $c_{s0}$ is the initial, uniform sound speed (proportional to the initial temperature) and $\tau_{cool}$ regulates the timescale of the simple effective cooling term, introduced to avoid an indefinite heating by viscous and resistive dissipation.

The time evolution of the magnetic field is done by uncurling the usual induction equation, using $\nabla\times \mathbf{A}=\mathbf{B}$, where $\mathbf{A}$ is the vector potential. For our system, the coupling between the shearing velocity field and the magnetic field reads \citep{Shearderivation1995}:
\begin{equation}
  \dfrac{\partial \mathbf{A}}{\partial t} = \mathbf{u}\times (\nabla \times \mathbf{A}) + \eta \nabla^2 \mathbf{A}~.
\label{Induction equation}
\end{equation}
The solenoidal condition, $\nabla \cdot \mathbf{B} = 0$, is therefore fulfilled at all times, avoiding any possible spurious growth coming from divergence cleaning procedures.

\subsection{Forcing}
\label{Subsec:forcing}

We impose an irrotational forcing, as a gradient of a Gaussian potential function:
\begin{equation}
  \mathbf{f}(\mathbf{x},t) = \nabla \phi(\mathbf{x},t) = \nabla \big( K \exp \{ -(\mathbf{x} - \mathbf{x_f}(t))^2 /R^2 \} \big)~,
\end{equation}
where $\mathbf{x}=(x,y,z)$ is the position vector, $\mathbf{x}_f(t)$ is the random position corresponding to the center of the expansion wave, $R$ is the radius of the Gaussian, and $N$ is the normalization factor. We will make use of $k_f$ as the wavenumber corresponding to the scale of the adopted forcing:
\begin{equation*}
k_f=\frac{2}{R}~. 
\end{equation*}
The time dependence of the forcing enters in the coherence time $\Delta t$ of the expansion waves (i.e., the time interval after which we consider a new random position, $\mathbf{x}_f$). We consider two different cases: in the first one $\mathbf{x}_f$ changes at every time-step (which is adaptive), $\Delta t=\delta t$, while in the second case we define an interval forcing time $\Delta t > \delta t$. We chose the normalization factor to be:
\begin{equation*}
  K =\phi_0 \sqrt{c_{s0} R /\Delta t}~,
\end{equation*}
where $\phi_0$ controls the forcing amplitude and has dimensions of the velocity squared.

After an initial transitory phase, the simulations reach a stationary state, over the course of which, the main average quantities maintain a saturated value. In particular, we will look at the root mean square of the velocity, $u_{rms}$. In turn, this is used to define the fundamental timescale of our problem, that we will call turnover time, as 
\begin{equation}
t_{turn} = (k_f u_{rms})^{-1}~.
\end{equation}
The turnover time can be understood as the average time for the fluid to cross en explosion width. 

The root mean square values of velocity $u_{rms}$ and vorticity $\omega_{rms}$ (see Sec. \ref{Subsec: vorticity} for its definition) are used to define the following dimensionless numbers:
\begin{eqnarray}
&&    \text{Re} = \frac{u_{rms}}{\nu k_f}~, \quad \quad \text{Rm} = \frac{u_{rms}}{\eta k_f}~, \quad \quad \text{Re$_\omega$} = \frac{\omega_{rms}}{\nu k_f^2}~,\\
&& \text{Ma}=u_{rms}/c_s ~, \quad \quad  \text{Pm}=\nu/\eta~,
\end{eqnarray}
which are the Reynolds number, magnetic Reynolds number, vorticity Reynolds number, Mach number, and magnetic Prandtl number, respectively. Analogously, we shall consider the maximum values of the velocity, $u_{max}$, to define, for example: 
\begin{eqnarray}
&& \text{Re}_M = \frac{u_{max}}{\nu k_f}~, \quad \quad \text{Ma}_M=u_{max}/c_s ~.
\end{eqnarray}
Regarding the magnetic fields, we shall consider the root-mean square, $b_{rms}$, closely related to the magnetic energy density.

\subsection{Shear}
\label{Subsec:shear}

In order to maintain a given shear velocity background along the $y$-direction $u_y^S$, we use a forcing term in the $y$-component of the velocity: $\mathbf{f}_s = \tau (u_y^S - u_y) \mathbf{\hat{y}}~$, where $\tau$ is the timescale of the forcing (which we keep unity in all cases).

We choose to apply a sinusoidal background flow for the shearing term:
\begin{equation}
    u_y^S = A \cos (k z)~.
    \label{Sinusoidal shear profile}
\end{equation}
As z ranges from $-\pi$ to $\pi$ (see Sec. \ref{Subsec: boundary}), we have set $k = 1$ which allows simple periodic boundary conditions in the three directions. This profile is similar to what is used by \cite{Skoutnev2022MNRAS} for studying dynamo in stellar interiors with a non-helical forcing, and was employed by \cite{Kayplaetal2009,Kayplaetal2009sinusoidal} in the context of stratified convective medium. 

We discarded the use of a more standard linear shear term $u_y^S=Sz$, since the implemented shearing boundary conditions were giving a spurious growth of vorticity at the boundaries (see Appendix \ref{Linear shear} for a more detailed explanation).

\subsection{Vorticity}
\label{Subsec: vorticity}

The vorticity is defined as the curl of the velocity field ($\mathbf{\omega} = \nabla \times \mathbf{u}$) and it quantifies the turbulent motions generated in fluid flows. Vortical motions are directly connected to the turbulent dynamo action, which is the mechanism by which our model may grow or maintain the magnetic field. The evolution of vorticity can be obtained by taking the curl of the total velocity evolution equation ($\partial \mathbf{u}/\partial t$):
\begin{equation}
\label{Vorticity evolution}
\begin{aligned}
  \frac{\partial \boldsymbol{\omega}}{\partial t} = \boldsymbol{\nabla} \times (\mathbf{u} \times \boldsymbol{\omega}) +  \nabla \times \mathbf{F}_{visc} - 2 \nabla \times \boldsymbol{\Omega} \times \mathbf{u} + \frac{\nabla \rho \times \nabla p}{\rho^2}+ \\  + \nabla \times \frac{\mathbf{J} \times \mathbf{B}}{\rho}  + \nabla  \times \mathbf{f}_s~.
\end{aligned}
\end{equation}
This equation is not directly evolved in our calculations and it serves only to understand the production mechanisms of vorticity, as each of the different terms of the RHS can generate or destroy it. We note that the forcing is irrotational by construction ($\mathbf{\nabla} \times \mathbf{f} = 0$), thus it does not appear in Eq. (\ref{Vorticity evolution}) and it cannot directly generate vorticity. However, the forced flow could help generate vorticity by the first two right-hand side terms: the first term (associated to turbulent motions) and the viscous terms. These two terms are quite analogous to the induction equation, hinting that turbulence amplification should be as difficult to achieve as dynamo action, in the absence of any other forcing. \cite{MeeBrandenburg2006} proved that this forcing produced no measurable vorticity: the amount of vorticity produced showed a decreasing dependence on numerical resolution, hence they concluded its nature was numerical.

The other vorticity sources appearing in the RHS are: rotation, baroclinic (zero for the barotropic case), Lorentz and shear terms, respectively. We expect the amount of vorticity generated by rotation to be proportional to the Coriolis number Co = $2 \Omega t_{\Omega}$, for a fixed viscosity and a rotation that is not too rapid. Also, the $t_{\Omega}$ here is a typical timescale of the system that we can assume $t_\Omega \approx t_{turn}$. The contribution of the rotation term to the vorticity, $\omega_{rms}$, can be then written as:

\begin{eqnarray}
\label{Vorticity rotation}
 && \frac{\partial \boldsymbol{\omega}}{\partial t} = ... - 2 \nabla \times \boldsymbol{\Omega} \times \mathbf{u} \nonumber \\
 && \rightarrow{} \omega_{rms} \approx 2~\Omega~ t_{turn}~u_{rms}~k_f \rightarrow{} \frac{k_{\omega}}{k_f} \approx 2 ~\Omega~t_{turn}~,
\end{eqnarray}
where we have defined: 
\begin{equation}
k_{\omega}=\frac{\omega_{rms}}{u_{rms}}~,
\end{equation}
as a proxy of how much of the available kinetic energy is in the form of vortical flows and provides an idea of the typical wavenumber of vortical structures. 

The shearing term contribution to Eq. (\ref{Vorticity evolution}) is:
\begin{eqnarray}
 && \frac{\partial \boldsymbol{\omega}}{\partial t} = ... +  \nabla  \times \left( - u_z \frac{\partial u^S_y}{\partial z} \mathbf{\hat{y}} \right)  \rightarrow{} \nonumber\\
 && \rightarrow{} \frac{\omega_{rms}}{t_{turn}} \approx  k_f~u_{rms}~k_f~u^S \rightarrow{}  \frac{k_{\omega}}{k_f} \approx A~t_{turn}~k_f~.
\end{eqnarray}
Finally, the baroclinic contribution $\nabla \rho \times \nabla p$ (misalignment between the pressure and density gradients) is proportional to $\nabla T \times \nabla s$ \citep[see Eqs. 12 and 13 in ][]{DelSordoBrandenburg2011}. To study the effect of such baroclinic term we can monitor the mean angle between these two gradients:
\begin{equation}
\text{ sin} \theta = \frac{\langle \nabla T \times \nabla s \rangle}{\langle \nabla T \rangle \langle \nabla s \rangle}~,\label{eq:sintheta}
\end{equation}
so that the evolution of vorticity can be seen as:
\begin{equation}
\label{Vorticity baroclinic}
  \frac{\partial \boldsymbol{\omega}}{\partial t} = ... -  \nabla T \times \nabla s  \quad \rightarrow{} \quad \omega_{rms} \approx t_{turn} (\nabla T)_{rms} (\nabla s)_{rms} \text{sin} \theta~.
\end{equation}

\subsection{Boundary and initial conditions}
\label{Subsec: boundary}

The simulation domain consists of a uniform, cubic grid mesh $[-\pi,\pi]^3$, with triply periodic boundary conditions. We consider resolutions varying from 128$^3$ up to 512$^3$, which are enough to assess our problem.

We adopted non-dimensional variables by measuring speed in units of the initial sound speed, $c_{s0}$, length in units of $1/k_1$ where $k_1$ is the smallest wave number in the periodic domain, implying that the non-dimensional size of the domain is $(2\pi)^3$.

As the initial conditions, pressure (and entropy and temperature in the baroclinic case), and density are set constant and with value 1 throughout the box. The flow is initially still, $\mathbf{v}=0$.

Finally, we set a weak initial magnetic field. Even though the initial flow is static, it is soon shaken by the expansion waves, so that the small initial magnetic seed does not influence the flow. For the initial magnetic topology, we consider either a uniform field in one given direction, or a spatially random field. The latter is assigned by picking, at each point, uncorrelated random values to the three components of the initial magnetic potential. The corresponding initial magnetic energy spectrum follows a $k^4$ power law, as reported by \cite{MeeBrandenburg2006}. 


\section{Results} \label{Results}

\subsection{Explored parameters}

We ran a series of different numerical models to explore the role of each term in the RHS of Eq. (\ref{Equation_of_motion}). The forcing and viscous terms are always included in all of our models, while we separately consider  the activation of either rotation or shear. The forcing amplitudes are high enough to create transonic flows, but we do not investigate supersonic flow. We have restrained ourselves from the usage of shock viscosities to avoid the introduction of numerical artifacts and to concentrate on the effect of a uniform viscosity. Most simulations have Pm=1 (and therefore Rm=Re) up to values of a few hundreds. In some runs, we went to Pm up to 100.

We start from some basic hydrodynamic (HD) simulations, similar to \cite{DelSordoBrandenburg2011}. We then expand them to magnetized cases, exploring broadly the parameter space. Unless otherwise specified, the expansion width has been chosen to be $R=0.2$ (thus, $k_f$ is 10).

In Tables \ref{table HD}, \ref{table barotropic}, \ref{table baroclinic}, \ref{table tau}, \ref{table high forcing}, and \ref{table sinusoidal shear}, we list the input parameters of the performed simulations. The tagging names starting with "H" indicate HD runs, those starting with "M" indicate MHD runs. By default, we employed the barotropic EoS, while "B" identifies the baroclinic cases. The first number stands for the value of $\Omega$. By default, we employ $\Delta t = 0.02$, $s$ in the end of the name identifies $\Delta t = 1$, while $c$ means $\Delta t = \delta t$. Unless stated otherwise (by the suffix 128, 512), the resolution is 256$^3$. The width ($R$) and constant ($\phi_0$) of the Gaussian are marked with the number following W and F in the run names, and, if not stated, they are assumed to be 0.2 and 1, respectively. The cooling term of the baroclininc cases is used only in the runs listed in Table \ref{table tau}. The initial magnetic field by default takes the form of random values of the potential vector component, unless a background uniform field in a given $i$ direction  is assigned (Bi). Finally, we  use "high" or "low" to indicate the changes in the viscous terms.

\begin{table}[t]
\centering
\scriptsize
\caption{HD simulations with different rigid rotation rates, and no shear. The quantities in the five columns are those defined in  Sec. \ref{sec:mhdeq} and \ref{Subsec:forcing}.}
\begin{tabular}{@{}cccccccc@{}}
\hline  \hline \\[-2.0ex]
256$^3$       & $\nu$   & $\chi$   & $\Omega$ & $\phi_0$ & $\Delta$t &  R   \\ \hline \\[-2.0ex]
H\_0s         & 2$\cdot$10$^{-4}$ & -           & 0        & 1       & 1       & 0.2           \\
H\_2s         & 2$\cdot$10$^{-4}$ & -           & 2        & 1       & 1       & 0.2           \\
H\_0          & 2$\cdot$10$^{-4}$ & -           & 0        & 1       & 0.02      & 0.2           \\
H\_2          & 2$\cdot$10$^{-4}$ & -           & 2        & 1       & 0.02      & 0.2           \\
H\_0c         & 2$\cdot$10$^{-4}$ & -           & 0        & 1       & $\delta t$         & 0.2           \\
H\_2c         & 2$\cdot$10$^{-4}$ & -           & 2        & 1       & $\delta t$        & 0.2           \\ 
H\_0cW1       & 2$\cdot$10$^{-3}$ & -           & 0        & 1       & $\delta t$         & 1           \\
H\_2cW1       & 2$\cdot$10$^{-3}$ & -           & 2        & 1       & $\delta t$         & 1           \\ \hline \\[-2.0ex]
HB\_0         & 2$\cdot$10$^{-4}$ &  2$\cdot$10$^{-4}$       & 0        & 1       & 0.02      & 0.2           \\
HB\_2         & 2$\cdot$10$^{-4}$ &  2$\cdot$10$^{-4}$       & 2        & 1       & 0.02      & 0.2           \\ \hline \\[-2.0ex] \hline
\end{tabular}
\label{table HD}
\end{table}

\begin{figure}[t]
\centering
\includegraphics[width=\hsize]{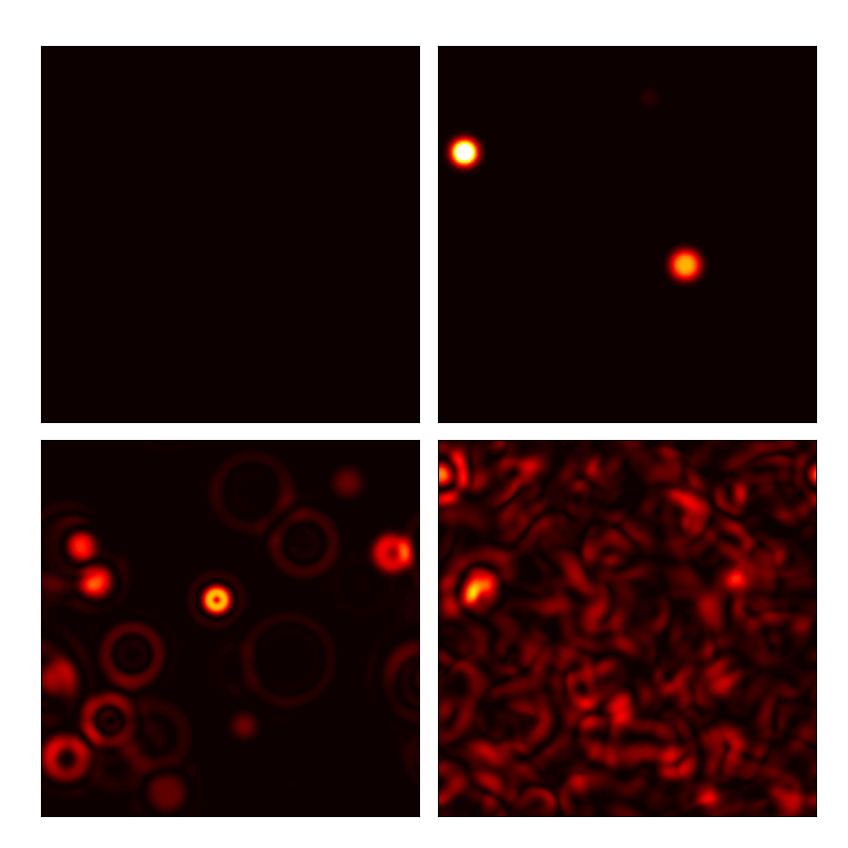}
    \caption{Slice in the $xy$-plane showing the value of $u^2$ at $t/t_{turn}= 0, 0.01$ (top), 0.1, and 1 (bottom) for the non-rotating model H\_0cW1 of Table~\ref{table HD}.}
    \label{slices_u2}
\end{figure}

\begin{table*}[ht]
\centering
\scriptsize
\caption{Input and diagnostics of the MHD barotropic runs.}
\begin{tabular}{@{}cccccccccccccccc@{}}
\hline \hline \\[-2.0ex]
128$^3$         & $\nu$    & $\eta$   &  B$_0$(G)  & $\Omega$ & $\phi_0$ & $\Delta$t & R  & t$_{turn}$      & k$_{\omega}$/k$_{f}$      & Re (Rm)   & Re$_{\omega}$       &  u$_{rot}$/u$_{tot}$            \\ \hline \\[-2.0ex]
M\_0\_128     & 2$\cdot$10$^{-4}$ & 2$\cdot$10$^{-4}$    & 10$^{-6}$             & 0        & 1       & 0.02      & 0.2  & 1.32            & 0.04693                   & 38.0 & 1.78   &0.039    \\ 
\textcolor{blue}{M\_1\_128}    & 2$\cdot$10$^{-4}$ & 2$\cdot$10$^{-4}$    & 10$^{-6}$             & 1        & 1       & 0.02      & 0.2  & 1.12            & \textcolor{blue}{0.190}   & 44.8 & 8.53      & 0.131      \\
M\_2\_128     & 2$\cdot$10$^{-4}$ & 2$\cdot$10$^{-4}$    & 10$^{-6}$             & 2        & 1       & 0.02      & 0.2  & 0.84            & 0.385                     & 59.6 & 22.9     & 0.284      \\
\textcolor{blue}{M\_3\_128}     & 2$\cdot$10$^{-4}$ & 2$\cdot$10$^{-4}$    & 10$^{-6}$             & 3        & 1       & 0.02      & 0.2  & 0.77            & \textcolor{blue}{0.543}   & 64.6 & 35.1     & 0.417     \\
M\_4\_128     & 2$\cdot$10$^{-4}$ & 2$\cdot$10$^{-4}$    & 10$^{-6}$             & 4        & 1       & 0.02      & 0.2   & 0.76            & 0.606                     & 65.6 & 39.7    & 0.481     \\
M\_5\_128     & 2$\cdot$10$^{-4}$ & 2$\cdot$10$^{-4}$    & 10$^{-6}$             & 5        & 1       & 0.02      & 0.2  & 0.78            & 0.642                     & 64.0 & 41.1  & 0.524        \\\hline \\[-2.0ex]
M\_0\_128\_Pr & 2$\cdot$10$^{-3}$ & 2$\cdot$10$^{-5}$    & 10$^{-6}$             & 0        & 1       & 0.02      & 0.2  &  1.93  &  0.00565  &   2.6 (259.4) &  0.01       \\ 
M\_2\_128\_Pr & 2$\cdot$10$^{-3}$ & 2$\cdot$10$^{-5}$    & 10$^{-6}$             & 2        & 1       & 0.02      & 0.2  &    1.86  &   0.29462  &  2.7 (269.2) &   0.79       \\ \hline \hline \\[-2.0ex]
256$^3$         & $\nu$    & $\eta$   &  B$_0$(G)   & $\Omega$ & $\phi_0$ & $\Delta$t & R   & t$_{turn}$      & k$_{\omega}$/k$_{f}$      & Re (Rm)   & Re$_{\omega}$         &  u$_{rot}$/u$_{tot}$           \\ \hline \\[-2.0ex]
M\_0          & 2$\cdot$10$^{-4}$ & 2$\cdot$10$^{-4}$    & 10$^{-6}$            & 0        & 1       & 0.02      & 0.2   & 1.42            & 0.01820                   & 35.3 & 0.64   & 0.014      \\
M\_0s          & 2$\cdot$10$^{-4}$ & 2$\cdot$10$^{-4}$    & 10$^{-6}$           & 0        & 1       & 1         & 0.2   & 4.32       & 0.00061                   & 11.6  & 0.007            \\
M\_0c         & 2$\cdot$10$^{-4}$ & 2$\cdot$10$^{-4}$    & 10$^{-6}$            & 0        & 1       & $\delta t$         & 0.2   & 1.46       & 0.01609              & 34.3  & 0.55          \\
M\_0low$\dagger$       & 2$\cdot$10$^{-5}$ & 2$\cdot$10$^{-5}$    & 10$^{-6}$            & 0        & 1       & 0.02      & 0.2      & 3.05       & 0.5913              & 163.8  & 97.03        \\
M\_0low\_F2$\dagger$   & 2$\cdot$10$^{-5}$ & 2$\cdot$10$^{-5}$    & 10$^{-6}$            & 0        & 2       & 0.02      & 0.2      & 1.03       & 1.4231              & 486.9  & 708.34        \\
M\_0lowc      & 2$\cdot$10$^{-5}$ & 2$\cdot$10$^{-5}$    & 10$^{-6}$            & 0        & 1       & $\delta t$         & 0.2   &   1.38   & 0.08582   & 362.6  &  31.12        \\
M\_0highcF10     & 2$\cdot$10$^{-2}$ & 2$\cdot$10$^{-2}$    & 10$^{-6}$            & 0        & 10       & $\delta t$         & 0.2       & 0.61       & 0.03443             & 0.81  & 0.028     \\ \hline \\[-2.0ex]
\textcolor{blue}{M\_2}          & 2$\cdot$10$^{-4}$ & 2$\cdot$10$^{-4}$    & 10$^{-6}$            & 2        & 1       & 0.02      & 0.2   & 0.84            & \textcolor{blue}{0.3809}   & 59.2 & 22.55    &  0.292      \\
M\_2c         & 2$\cdot$10$^{-4}$ & 2$\cdot$10$^{-4}$    & 10$^{-6}$            & 2        & 1       & $\delta t$         & 0.2  &   0.85  & 0.38179  & 58.5  & 22.35         \\
M\_2W0.5$\dagger$      & 2$\cdot$10$^{-4}$ & 2$\cdot$10$^{-4}$    & 10$^{-6}$            & 2        & 1       & $\delta t$         & 0.5   &   0.29  & 0.64222  & 171.7 &  110.24       \\
M\_2low$\dagger$       & 2$\cdot$10$^{-5}$ & 2$\cdot$10$^{-5}$    & 10$^{-6}$            & 2        & 1       & 0.02       & 0.2  &  2.91   & 1.60296   & 171.8  &  276.5  \\ \hline \\[-2.0ex]
M\_0W0.1        & 2$\cdot$10$^{-2}$     & 2$\cdot$10$^{-2}$    & 10$^{-9}$    & 0      & 1       & 0.02     & 0.1    & 13.25  &  0.00073  &  0.0094   &  0.00001       \\
M\_0W0.2        & 2$\cdot$10$^{-2}$     & 2$\cdot$10$^{-2}$    & 10$^{-9}$    & 0      & 1       & 0.02     & 0.2    & 6.51   &  0.00348   &  0.0077   &  0.00027   \\
M\_0W0.5        & 2$\cdot$10$^{-2}$     & 2$\cdot$10$^{-2}$    & 10$^{-9}$    & 0      & 1       & 0.02     & 0.5    & 2.79   &  0.00177   &  1.12  &   0.0198    \\
M\_0W1          & 2$\cdot$10$^{-2}$     & 2$\cdot$10$^{-2}$    & 10$^{-9}$    & 0      & 1       & 0.02      & 1.0   & 1.96   &  0.05571   &  6.37 &   0.355     \\ \hline \hline \\[-2.0ex]
256$^3$         & $\nu$    & $\eta$   &  B$_0$(U)   & $\Omega$ & $\phi_0$ & $\Delta$t & R     & t$_{turn}$      & k$_{\omega}$/k$_{f}$      & Re (Rm)  & Re$_{\omega}$        &  u$_{rot}$/u$_{tot}$          \\ \hline \\[-2.0ex]
M\_0B        & 2$\cdot$10$^{-4}$ & 2$\cdot$10$^{-4}$    & 10$^{-2}$            & 0        & 1       & 0.02      & 0.2    & 1.42            & 0.01820                   & 35.3 & 0.64            \\
M\_2Bxs        & 2$\cdot$10$^{-4}$ & 2$\cdot$10$^{-4}$    & 10$^{-6}$            & 2        & 1       & 1         & 0.2  &  3.93  & 0.39751  & 12.7  & 5.05          \\
M\_2Bx        & 2$\cdot$10$^{-4}$ & 2$\cdot$10$^{-4}$    & 10$^{-2}$            & 2        & 1       & 0.02      & 0.2   &   0.85  & 0.3809  & 59.2 &  22.54          \\
M\_2By        & 2$\cdot$10$^{-4}$ & 2$\cdot$10$^{-4}$    & 10$^{-2}$            & 2        & 1       & 0.02      & 0.2        &   0.85  & 0.3808  & 59.2 &  22.53      \\
M\_2Bz        & 2$\cdot$10$^{-4}$ & 2$\cdot$10$^{-4}$    & 10$^{-2}$            & 2        & 1       & 0.02      & 0.2      &   0.84  & 0.3821  & 59.2 &  22.63        \\ \hline \hline \\[-2.0ex]
512$^3$         & $\nu$    & $\eta$   &  B$_0$(G)   & $\Omega$ & $\phi_0$ & $\Delta$t & R      & t$_{turn}$      & k$_{\omega}$/k$_{f}$      & Re (Rm)   & Re$_{\omega}$         &  u$_{rot}$/u$_{tot}$          \\ \hline \\[-2.0ex]
\textcolor{blue}{M\_0\_512}     & 2$\cdot$10$^{-4}$ & 2$\cdot$10$^{-4}$    & 10$^{-6}$           & 0        & 1       & 0.02      & 0.2 & 1.47            & \textcolor{blue}{0.00566} & 34.0 & 0.19      &  0.012      \\ \hline \hline
\end{tabular}
\label{table barotropic}
\tablefoot{We indicate the initial amplitude of the magnetic field $B_0$ along with ('G') for random values for the potential vector components or ('U$_i$') for a uniform distribution in a given direction $i=\{x,y,z\}$. The included diagnostic magnitudes are: turnover time, vorticity proxy $k_\omega/k_f$, Re (and Rm if Pm $\neq 1$), and Re$_\omega$. The last column corresponds to the rotational flow contribution obtained using the Helmholtz decomposition only for some specific runs. The blue highlighted values are used to perform the linear fit $k_\omega/k_f(\Omega)$ discussed in the text. The four simulations marked with $\dagger$ became numerically unstable before reaching a fully steady saturated state: we still indicate their diagnostics, even if it is not completely comparable to the others.}
\end{table*}

\begin{table*}[ht]
\centering
\scriptsize
\caption{Input and diagnostics of the MHD baroclinic runs.}
\begin{tabular}{@{}cccccccccccccc@{}}
\hline \hline \\[-2.0ex]
128$^3$       & $\nu$    & $\chi$   & $\eta$   & B$_0$(G)    & $\Omega$ & $\phi_0$ & $\Delta$t & R & t$_{turn}$      & k$_{\omega}$/k$_{f}$      & Re (Rm)   & Re$_{\omega}$     \\ \hline \\[-2.0ex]
MB\_0\_128    & 2$\cdot$10$^{-4}$    & 2$\cdot$10$^{-4}$     & 2$\cdot$10$^{-4}$    & 10$^{-6}$          & 0        & 1       & 0.02      & 0.2   & 1.59 & 0.0086 &  31.5 & 0.27         \\
MB\_2\_128    & 2$\cdot$10$^{-4}$    & 2$\cdot$10$^{-4}$     & 2$\cdot$10$^{-4}$    & 10$^{-6}$          & 2        & 1       & 0.02      & 0.2    & 1.20 & 0.195  &  41.8 & 8.19          \\ \hline \\[-2.0ex]
MB\_0\_128\_Pr & 2$\cdot$10$^{-3}$    & 2$\cdot$10$^{-3}$     & 2$\cdot$10$^{-5}$    & 10$^{-6}$          & 0        & 1       & 0.02      & 0.2  &   2.50 &  0.00319  & 2.002 (200.2)  & 0.0064       \\
MB\_2\_128\_Pr & 2$\cdot$10$^{-3}$    & 2$\cdot$10$^{-3}$     & 2$\cdot$10$^{-5}$    & 10$^{-6}$          & 2        & 1       & 0.02      & 0.2  & 2.39  & 0.21101  & 2.095 (209.5)  & 0.44
      \\ \hline \hline \\[-2.0ex]
256$^3$        & $\nu$    & $\chi$   & $\eta$   & B$_0$(G)    & $\Omega$ & $\phi_0$ & $\Delta$t & R  & t$_{turn}$      & k$_{\omega}$/k$_{f}$      & Re (Rm)  & Re$_{\omega}$      \\ \hline \\[-2.0ex]
MB\_0          & 2$\cdot$10$^{-4}$    & 2$\cdot$10$^{-4}$ & 2$\cdot$10$^{-4}$    & 10$^{-6}$          & 0        & 1       & 0.02      & 0.2    & 1.65 & 0.0878 &  30.3 & 0.27      \\
MB\_0c         & 2$\cdot$10$^{-4}$    & 2$\cdot$10$^{-4}$ & 2$\cdot$10$^{-4}$    & 10$^{-6}$          & 0        & 1       & $\delta t$         & 0.2    & 1.65 & 0.00876 &  30.3 & 0.27         \\
MB\_0highcF10  & 2$\cdot$10$^{-2}$    & 2$\cdot$10$^{-4}$ & 2$\cdot$10$^{-2}$    & 10$^{-6}$          & 0        & 10       & $\delta t$         & 0.2    &   0.64  & 0.01376 &  0.78 &  0.0108     \\
MB\_0lowc     & 5$\cdot$10$^{-5}$    & 2$\cdot$10$^{-4}$ & 5$\cdot$10$^{-5}$    & 10$^{-6}$          & 0        & 1       & $\delta t$         & 0.2      & 1.61 & 0.02876 & 124.1 & 3.57     \\ 
MB\_0low      & 5$\cdot$10$^{-5}$    & 2$\cdot$10$^{-4}$ & 5$\cdot$10$^{-5}$    & 10$^{-6}$          & 0        & 1       & 0.02         & 0.2      & 1.61 & 0.03046 & 124.4 & 3.79      \\
MB\_0low2c      & 2$\cdot$10$^{-5}$    & 2$\cdot$10$^{-4}$ & 2$\cdot$10$^{-5}$    & 10$^{-6}$          & 0        & 1       & $\delta t$         & 0.2      & 1.60 & 0.0434 & 313.1 & 13.58      \\\hline \\[-2.0ex]
MB\_2          & 2$\cdot$10$^{-4}$    & 2$\cdot$10$^{-4}$ & 2$\cdot$10$^{-4}$    & 10$^{-6}$          & 2        & 1       & 0.02      & 0.2    & 1.08 & 0.340  &  46.5 & 15.83        \\
MB\_2c         & 2$\cdot$10$^{-4}$    & 2$\cdot$10$^{-4}$ & 2$\cdot$10$^{-4}$    & 10$^{-6}$          & 2        & 1       & $\delta t$         & 0.2    & 1.06 & 0.343  &  47.0 & 16.12       \\
MB\_2low$\dagger$      & 2$\cdot$10$^{-5}$    & 2$\cdot$10$^{-4}$ & 2$\cdot$10$^{-5}$    & 10$^{-6}$          & 2        & 1       & 0.02      & 0.2   & 0.96  &  0.6172  &  522.2  &  323.0       \\ \hline  \\[-2.0ex]
MB\_0W0.1       & 2$\cdot$10$^{-2}$     & 2$\cdot$10$^{-2}$     & 2$\cdot$10$^{-2}$    & 10$^{-9}$   & 0        & 1       & 0.02      & 0.1   & 15.31  & 0.00076  & 0.0082 &  0.00001      \\
MB\_0W0.2       & 2$\cdot$10$^{-2}$     & 2$\cdot$10$^{-2}$     & 2$\cdot$10$^{-2}$    & 10$^{-9}$   & 0        & 1       & 0.02      & 0.2   & 7.84 & 0.00419 & 0.0638 & 0.00027      \\
MB\_0W0.5       & 2$\cdot$10$^{-2}$     & 2$\cdot$10$^{-2}$     & 2$\cdot$10$^{-2}$    & 10$^{-9}$   & 0        & 1       & 0.02      & 0.5   & 1.36 & 0.01923 & 0.9209 & 0.01772            \\
MB\_0W1         & 2$\cdot$10$^{-2}$     & 2$\cdot$10$^{-2}$     & 2$\cdot$10$^{-2}$    & 10$^{-9}$   & 0        & 1       & 0.02      & 1.0   & 0.46 & 0.06042 & 5.399 & 0.3263       \\\hline \hline \\[-2.0ex]
256$^3$        & $\nu$    & $\chi$   & $\eta$   & B$_0$(U)    & $\Omega$ & $\phi_0$ & $\Delta$t & R   & t$_{turn}$      & k$_{\omega}$/k$_{f}$      & Re (Rm)  & Re$_{\omega}$     \\ \hline \\[-2.0ex]
MB\_0B        & 2$\cdot$10$^{-4}$    & 2$\cdot$10$^{-4}$ & 2$\cdot$10$^{-4}$    & 10$^{-2}$         & 0        & 1       & 0.02      & 0.2   &   1.65 &  0.00898 &  30.3  & 0.27     \\
MB\_2Bx       & 2$\cdot$10$^{-4}$    & 2$\cdot$10$^{-4}$ & 2$\cdot$10$^{-4}$    & 10$^{-2}$         & 2        & 1       & 0.02      & 0.2     &   1.05  & 0.30828  & 47.5  & 14.63     \\
MB\_2By       & 2$\cdot$10$^{-4}$    & 2$\cdot$10$^{-4}$ & 2$\cdot$10$^{-4}$    & 10$^{-2}$         & 2        & 1       & 0.02      & 0.2    &   1.05   & 0.30744 &  47.7  & 14.65    \\
MB\_2Bz       & 2$\cdot$10$^{-4}$    & 2$\cdot$10$^{-4}$ & 2$\cdot$10$^{-4}$    & 10$^{-2}$         & 2        & 1       & 0.02      & 0.2    &   1.05  & 0.30858 &  47.5  & 14.66    \\ \hline  \hline \\[-2.0ex]
512$^3$         & $\nu$   & $\chi$  & $\eta$   &  B$_0$(G)   & $\Omega$ & $\phi_0$ & $\Delta$t & R      & t$_{turn}$      & k$_{\omega}$/k$_{f}$      & Re (Rm)  & Re$_{\omega}$              \\ \hline \\[-2.0ex]
MB\_0\_512      & 2$\cdot$10$^{-4}$ & 2$\cdot$10$^{-4}$ & 2$\cdot$10$^{-4}$    & 10$^{-6}$           & 0        & 1       & 0.02      & 0.2 & 1.69            & 0.00717 & 29.6 & 0.21           \\ \hline \hline
\end{tabular}
\label{table baroclinic}
\tablefoot{Table is organized the same way as Table \ref{table barotropic}. The simulation marked with $\dagger$ became unstable before reaching a fully steady saturated state: we still indicate  its diagnostics, although they are not completely comparable to the others.}
\end{table*}

We monitor the volume-integrated energies and the above mentioned root mean square values of quantities. Additionally, we look at the spectral energy distribution of velocity (total or turbulent), vorticity, and magnetic field, as a function of the wavenumber, $k$ (which includes the factor 2$\pi$, see Appendix~\ref{Spectra}). Spectra are also averaged over different times to filter out fluctuations, which are quite noticeable on the smallest scales. To evaluate the rotationality of the flow we also make use the Helmholtz decomposition for the velocity field, as defined in Appendix \ref{Helmholtz decomposition}.

\begin{figure}[t]
\centering
\includegraphics[width=\hsize]{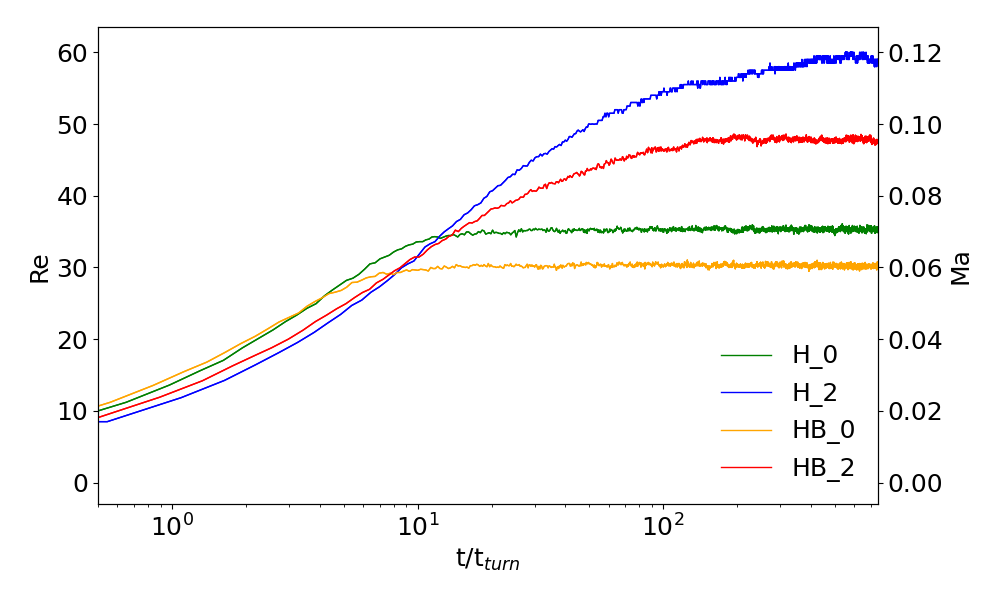}
    \caption{Comparison of Re and Ma evolution in the HD cases, considering or not the {\bf rigid} rotation ($\Omega=0$ or 2) and for different EoS (H vs. HB).}
    \label{Re_turnover}
\end{figure}

\begin{figure}[h]
\centering
\includegraphics[width=\hsize]{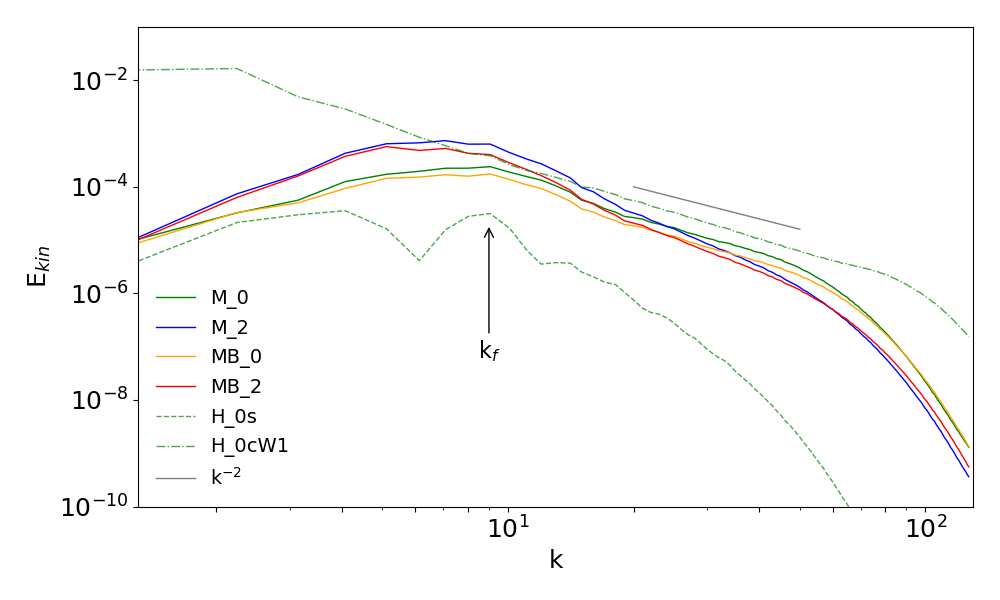}
\caption{Time-averaged kinetic spectra at saturation of some representative simulations of Tables \ref{table HD}, \ref{table barotropic}, and \ref{table baroclinic}. 
We notice how the results on large scales (small k) are very similar for all the models, excluding H\_0cW1 (dash-dotted green line), which has a forcing with $k_f=2$. The spectrum with lowest power is H\_0s (dashed green line), which has $\Delta t=1$. Models marked with other colors explore variation of $\Omega$ and EoS. In all these models the magnetic field decays rapidly and kinetic spectra coincide with the corresponding HD runs of Fig. \ref{Re_turnover}.}
\label{Spectra rotations}
\end{figure}

\subsection{Effects of forcing and rigid rotation on the flow}

Table~\ref{table HD} lists some examples of HD runs, without magnetic field. To visualize the general behavior at early times, Fig. \ref{slices_u2} shows the values of $u^2$ in the $xy$-plane for one of them, the representative model H\_0cW1. Rapidly, the imprints of the most recent expansion waves are visible on top of the turbulent state. Such an evolution leads to a homogeneous turbulent flow, which becomes stationary when the dissipating forces counterbalance the forced energy injection.

Such features are seen also in the MHD cases without shear, for which Tables \ref{table barotropic} and \ref{table baroclinic} list the barotropic and baroclinic configurations, respectively, considering or not rigid rotation. In these same tables, we provide the diagnostic magnitudes. As a first comparison, in Fig. \ref{Re_turnover} we can see how different representative simulations (with or without rotation, and for the two EoS) saturate their Re and Ma values within $\sim 1000~t_{turn}$. A faster rotation increases the final saturation value for the velocities, but the growth takes places at a similar rate, so simulations with faster rotation take longer to saturate. The ideal EoS (i.e., baroclinic) cases show a lower saturation value than the barotropic one. We interpret this result as a consequence of the presence of the extra dissipating terms in the energy equation. The maximum values for Re and Ma for these simulations, $\text{Re}_M$ and  $\text{Ma}_M$, oscillate much more (but with values that are similar for each), namely, ranging from 150 to 300, and 0.3 to 0.6, respectively. These values slightly increase with the rotation rate, and again are smaller in the baroclinic case.

Figure \ref{Spectra rotations} shows the comparison between kinetic spectra at saturation for different EoS and $\Omega$. In general, for all cases kinetic spectra peak around the value $k_f$ (simulations M\_0s and M\_0 having $R = 0.2$, show the bump for $k \sim 5-10$, while H\_0cW1 accumulates energy at the largest scales, since $R=1$). This is in agreement with the characteristic forcing wavenumber $k_f$ of the adopted forcing, which in the Fourier space is $\propto \exp(-k^2/k_f^2)$. The inertial part of the spectra shows a $\sim k^{-2}$ slope. This irrotational forcing does not lead to any vorticity production, which might help explain the difference with respect to the isotropic turbulent slope of $k^{-5/3}$. These general spectral features (peak and slope) are compatible with those found by \cite{MeeBrandenburg2006} and \cite{DelSordoBrandenburg2011}.

Rotation tends to inhibit smaller scales and promotes the accumulation of kinetic energy at the largest scales, also slightly displacing the spectra peak to the left. This results in a steeper slope in the inertial range. A similar (but less noticeable) effect is seen in the baroclinicity, as compared to the barotropic cases. This is in agreement with the lower total kinetic energy, due to the extra diffusive terms.

To quantitatively measure  the dependence of vorticity with rotation, we compared the saturation values of Re$_{\omega}$ (see Table \ref{table barotropic}). The saturation value increases with $\Omega$, but reaches a maximum around $\Omega$ = 5, from which rising rotation does not lead to more vorticity. This translates into a saturation of Rossby number, $\mathrm{Ro} := u_{rms} k_f/2 \Omega$, of the order of $u_{rms}$. The low amount of vorticity appearing with $\Omega$ = 0 appears to be spurious, since it decreases when resolution is increased, as already reported by \cite{MeeBrandenburg2006}. Such spurious contribution is anyway much smaller than the physical one when the rotation is considered.

In order to see if the estimate in Eq.~(\ref{Vorticity rotation}) holds, we show in Table \ref{table barotropic} some physical quantities of these runs. We find a linear trend valid up to $\Omega \sim 3$, which we quantify by a trend between some simulations (marked in blue in Table \ref{table barotropic}) as: $k_{\omega}/k_f$ = (0.180$\pm$0.004) $\Omega$ + (0.010$\pm$0.009). The slope seems much smaller for $\Omega\gtrsim 4$ (but further simulations would be needed to quantify it).

Finally, the saturated values increase as we decrease $\Delta t$ from 1 (H\_0s) to 0.02 (H\_0). There is not much of a difference between $\Delta t$ being 0.02 or $\delta t$, since the latter is indeed on the order ${\cal O}(10^{-2})$. Overall, these results are compatible with the idea that this irrotational forcing alone does not produce any significant amount of vorticity.

\begin{figure}[t]
\centering
\includegraphics[width=\hsize]{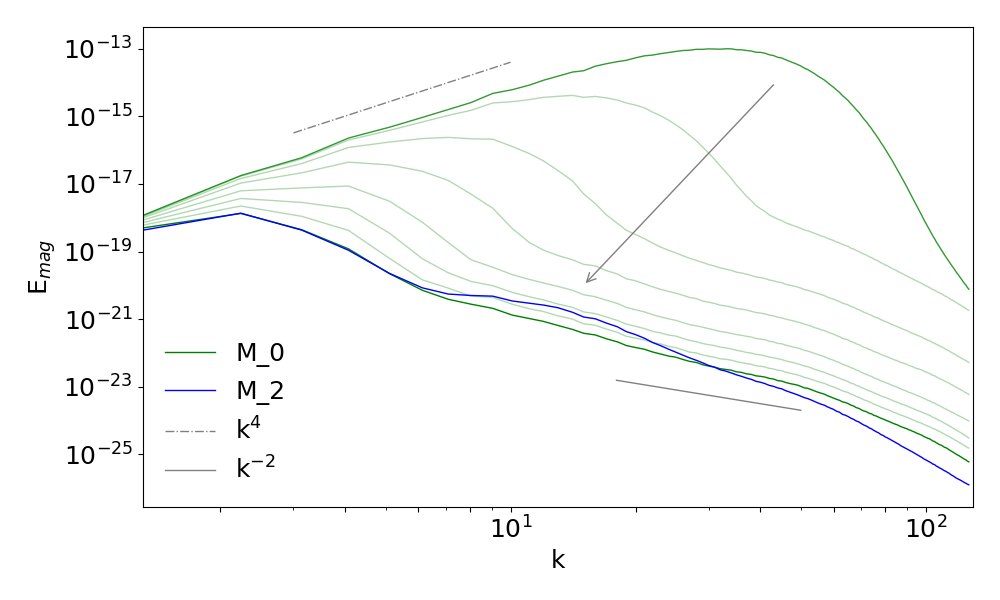}
    \caption{Magnetic spectra decay for the M\_0 and M\_2 models. The spectra evolution is only shown for M\_0, and the first corresponds to less than a turnover time when it still shows the initial slope of $k^4$. The arrow indicates the direction of temporal evolution. The last spectrum is approximately at turnover time 20 and it is shown for both models.}
    \label{Spectra decay rotations}
\end{figure}

\subsection{Absence of dynamo with rigid rotation only}

In all the models of Tables \ref{table barotropic} (barotropic) and \ref{table baroclinic} (baroclinic), the initial magnetic field quickly decays, with or without rotation. In other words, in spite of the vorticity growth induced by rotation, the system does not experience SSD. Indeed, there is no significant difference in the flow between these MHD models the  purely HD ones that correspond to them. 

An example of a decay of the magnetic spectrum is shown in Fig. \ref{Spectra decay rotations}, where it can be observed how the smallest scales decay faster and the $k^4$ slope rapidly changes.  Washing away first the smaller scales is natural, as both numerical and physical diffusivities have more influence on the very small scales. After some turnover times the inertial range of the magnetic spectra resemble (in some way) the kinetic ones. Rotation also tends to favor larger scales and increase the slope, although both slopes are steeper than in the kinetic spectra. There is a minimal interaction between the flow and the magnetic field through the first term of the induction, as expressed by Eq. (\ref{Induction equation}). 

The decaying evolution of b$_{rms}$, for the initial random values, were found to follow power laws. The parameter fits do not depend on $\Omega$ for our runs. With no forcing we obtain very similar decays with the same exponent, hence, finding no substantial difference between M\_0 and the corresponding simulation with no forcing. This leads us to think that the forced flow does not have a major destructive influence on the field, contrary to what \cite{MeeBrandenburg2006} noted, although it does not lead to any growth of the magnetic field either.

\begin{table}[t]
\centering
\scriptsize
\caption{MHD simulations with different values for $\tau_{cool}$.}
\begin{tabular}{@{}ccccccccccc@{}}
\hline \hline \\[-2.0ex]
256$^3$       & $\nu$    & $\chi$   & $\eta$   & B$_0$(G)    & $\tau_{cool}$ & $\phi_0$ & $\Delta$t & R  & Re (Rm) \\ \hline \\[-2.0ex]
MB\_t\_1.57    & 0.1     & 0.1      & 0.1     & 10$^{-6}$     & 1.5708      & 10       & 0.02      & 0.2    &   0.0980  \\ 
MB\_t\_3.14    & 0.1     & 0.1      & 0.1     & 10$^{-6}$     & 3.1415      & 10       & 0.02      & 0.2    &   0.0953  \\ 
MB\_t\_6.28    & 0.1     & 0.1      & 0.1     & 10$^{-6}$     & 6.2831      & 10       & 0.02      & 0.2    &   0.0870  \\ 
MB\_t\_31.4    & 0.1     & 0.1      & 0.1     & 10$^{-6}$     & 31.415      & 10       & 0.02      & 0.2    &   0.0759  \\ 
MB\_t\_314     & 0.1     & 0.1      & 0.1     & 10$^{-6}$     & 314.15      & 10       & 0.02      & 0.2    &   0.0726  \\
MB\_nt         & 0.1     & 0.1      & 0.1     & 10$^{-6}$     & -           & 10       & 0.02      & 0.2    &   0.0638  \\ \hline \hline 
\end{tabular}
\label{table tau}
\end{table}

\begin{table}
\centering
\scriptsize
\caption{Simulations involving higher forcing and diffusivities, using neither shear nor rotation.}
\begin{tabular}{@{}ccccccc@{}}
\hline \hline \\[-2.0ex]
                  & $\nu$    & $\chi$   & $\phi_0$ & $\Delta$t & R  & Re \\ \hline \\[-2.0ex]
M\_0highc\_F5    & 10$^{-3}$ & - & 5       & $\delta t$                        & 0.2   &  18.13      \\
M\_0highc\_F20   & 0.1 & - & 10      & $\delta t$                        & 0.2   &   0.1421      \\
MB\_0highc\_F50   & 0.1 & 0.1 & 50      & $\delta t$                        & 0.2   &  0.3467        \\ \hline \\[-2.0ex]
MB\_0high\_F50    & 0.1 & 0.1 & 50      & 0.02                     & 0.2    &   0.4319      \\
MB\_0high2\_F50   & 1    & 1    & 50      & 0.02                     & 0.2    &  0.0089      \\ \hline \\[-2.0ex]
MB\_0high\_F100   & 1 & 1 & 100     & 0.02                     & 0.2    &   0.0329     \\
MB\_0high\_F100\_W1$\dagger$  & 1    & 1    & 100     & 0.02                     & 1      &   0.3931    \\ \hline \\[-2.0ex]
MB\_0high\_F200   & 1 & 1 & 200     & 0.02                     & 0.2    &    0.0657    \\
MB\_0high\_F200\_W1$\dagger$  & 1    & 1    & 200     & 0.02                     & 1     &    0.8864     \\ \hline \\[-2.0ex]
MB\_0high\_F500   & 1 & 1 & 500     & 0.02                     & 0.2    &    0.0647    \\
MB\_0high\_F500\_W1$\dagger$  & 1    & 1    & 500     & 0.02                     & 1     &    3.4525     \\ \hline \hline
\end{tabular}
\label{table high forcing}
\end{table}

We have explored this for a large range of parameters, especially for the baroclinic cases, with rotation only. There was no SSD seen for any of the cases, despite thousands of runs at various turnover times. In particular, this conclusion is reached in each of these cases: (i) changing $R$, $\Delta t$ (Tables \ref{table barotropic} and \ref{table baroclinic}); (ii) using a simple uniform initial magnetic field, which experiences an early fast (exponential) decay, followed by a power law, with or without rotation; (iii) changing resolution; and (iv) changing $\tau_{cool}$ in the baroclinic cases (Table \ref{table tau}).

We found that by not including the cooling term in the entropy equation, the sine of the baroclinic angle, Eq.~(\ref{eq:sintheta}) takes peak values that are higher than one. The mean angle takes similar values of the ones found in \cite{Korpi1999ApJ...514L..99K}.
These results are in contrast with \cite{AchikanathFederrath2021} who found SSD in similar turbulent setups. In particular, even with a completely irrotational driving (implemented in the Fourier space) and without any rotation, they see that after $\sim 10^3 ~t_{turn}$ there is an increase of magnetic energy, which saturates at about 1/1000 of the kinetic one. To compare them, simulations MB\_0\_128 and MB\_2\_128 were run to approximately 10$^4$ $t_{turn}$. We also ran simulations with high values for the forcing amplitude (up to $\phi_0$=500, see Table \ref{table high forcing} for non-rotating cases), which required higher values of viscosity and thermal diffusivities, leading to overall lower values of Re. However, we still could not observe any growth in the magnetic field.

As we wanted to investigate as much as possible the parameter space looking for possible dynamo action, we tried to push the code capability to its limits by increasing the values of $\Omega$ or decreasing diffusivites. In both cases, we could see that a growth of the field could be found only in cases that quickly became numerical unstable. In these cases, we were using either $\Omega=10$ or diffusivities of the order of $\nu=10^{-6}$. As in these cases it was not possible to clearly establish growth and saturation phases, we did not consider them reliable and, hence, we did not include them in our analysis.

Summarizing, the irrotational forcing in combination with solid body rotation can produce vorticity. The chosen ideal EoS favors it more than the barotropic one. However, SSD is never activated, regardless of the rotation, EoS, and correlation length of the seed magnetic field, at least in the explored range of Rm up to a few hundred units. 

\subsection{Dynamo in the presence of a shear}

A further effect to be investigated is the shearing flow in combination with this irrotational forcing. In Table \ref{table sinusoidal shear} we report the numerical experiment performed with the sinusoidal shearing flow $\mathbf{u}^S$ introduced above in Sec. \ref{Subsec:shear}. Dynamo growth is always found, unless the shearing profile is rather weak (first three models in table \ref{table sinusoidal shear}), or the forcing acts on a rather small length-scale (case M\_0A020\_W0.10\_128), or if the forcing is not acting, meaning there is only the background flow (case M\_0A020\_F0\_128). In terms of Reynolds numbers, we tentatively indicate a threshold value of $\mathrm{Rm}_{crit}\sim$ 50, for the chosen setups.

\begin{figure}[t]
\centering
\includegraphics[width=\hsize]{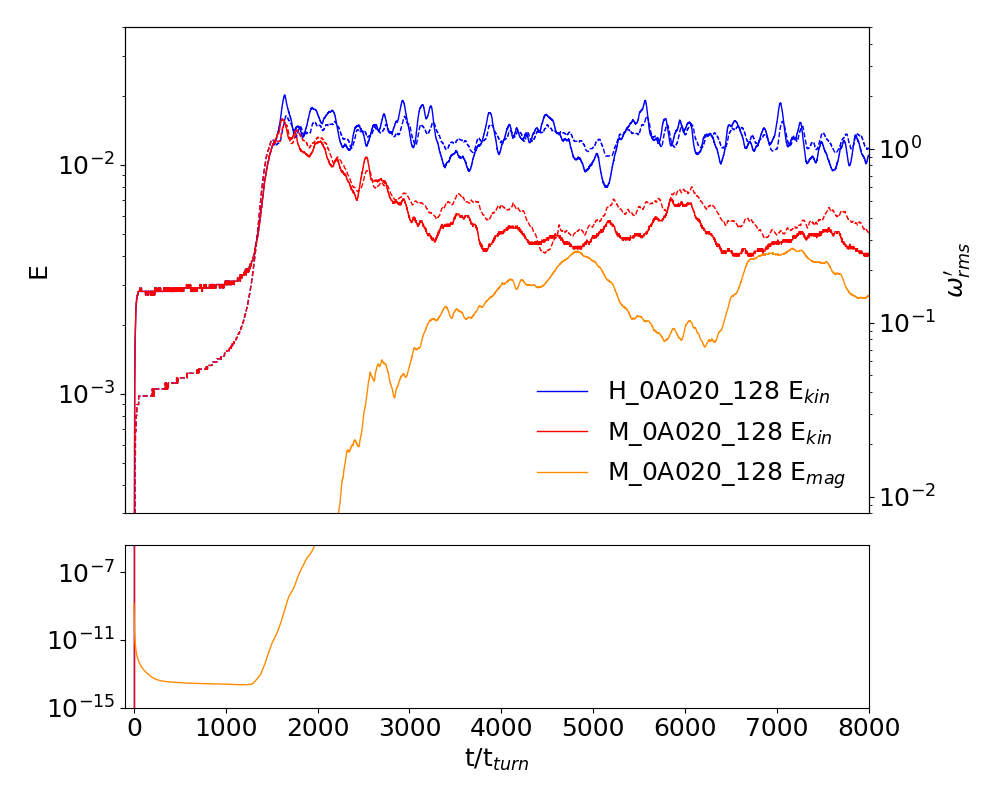}
\caption{Time evolution for the average turbulent kinetic energy density (solid blue and red lines), the average magnetic energy density (yellow line) and the root mean square of the turbulent vorticity $\omega'_{rms}$ (dashed blue and red lines) for one HD setup and for an MHD set up with the same physical parameters, M\_0A020\_128.}
\label{Hydro vs dynamo instability}
\end{figure}

The system evolves as illustrated in Fig. \ref{Hydro vs dynamo instability}. There is an initial growth of the vorticity that after a few turnover times seems to saturate, along with the value of $u_{rms}$. The small-scale magnetic field initially decays. For about 1000 turnover times, the system slowly evolves by slightly increasing the vorticity and keeping  both  $u_{rms}$ and $b_{rms}$ almost constant. We then observe a sudden growth of vorticity followed by that of the kinetic energy, and then also of the magnetic energy. This process occurs with a strong enough shear amplitude, for a given a set of diffusivities and forcing parameters. We understand it as a vorticity dynamo produced by a Kelvin-Helmholtz instability, which develops in the system after it is perturbed beyond a certain threshold.

\begin{table*}[ht]
\centering
\scriptsize
\caption{Input and Re of the MHD runs with sinusoidal shear.}
\begin{tabular}{cccccccccccc}
\hline \hline \\[-2.0ex]
128$^3$            & $\nu$    & $\chi$ & $\eta$   & B$_0$(G)   & $\Omega$ & $A$ & $\phi_0$ & $\Delta$t & R & Re & Dynamo  \\ \hline \\[-2.0ex]
M\_0A000\_128 & 2$\cdot$10$^{-4}$ & -      & 2$\cdot$10$^{-4}$ & 10$^{-6}$ & 0        & 0     & 1        & 0.02         & 0.2 &  38.0  & No \\
M\_0A002\_128 & 2$\cdot$10$^{-4}$ & -      & 2$\cdot$10$^{-4}$ & 10$^{-6}$ & 0        & 0.02  & 1        & 0.02         & 0.2 &  38.6  & No \\
M\_0A005\_128 & 2$\cdot$10$^{-4}$ & -      & 2$\cdot$10$^{-4}$ & 10$^{-6}$ & 0        & 0.05  & 1        & 0.02         & 0.2 &  41.9  & No \\
M\_0A010\_128 & 2$\cdot$10$^{-4}$ & -      & 2$\cdot$10$^{-4}$ & 10$^{-6}$ & 0        & 0.10  & 1        & 0.02         & 0.2 &  52.1  & Yes \\
M\_0A020\_128  & 2$\cdot$10$^{-4}$ & -      & 2$\cdot$10$^{-4}$ & 10$^{-6}$ & 0       & 0.20  & 1        & 0.02         & 0.2 &  80.2  & Yes *\\
M\_0A050\_128 & 2$\cdot$10$^{-4}$ & -      & 2$\cdot$10$^{-4}$ & 10$^{-6}$ & 0        & 0.50  & 1        & 0.02         & 0.2 &  180.3  & Yes \\
M\_0A100\_128 & 2$\cdot$10$^{-4}$ & -      & 2$\cdot$10$^{-4}$ & 10$^{-6}$ & 0        & 1.00  & 1        & 0.02         & 0.2 &  355.1  & Yes \\ \hline \\[-2.0ex]
H\_0A020\_128     & 2$\cdot$10$^{-4}$ & -    & -    & -    & 0        & 0.20  & 1        & 0.02         & 0.2 &  80.2  & - \\
H\_0A020\_F0\_128 & 2$\cdot$10$^{-4}$ & -    & -    & -    & 0        & 0.20  & 0        & -            & -   &  70.5  & - \\ 
H\_0A050\_128     & 2$\cdot$10$^{-4}$ & -    & -    & -    & 0        & 0.50  & 1        & 0.02         & 0.2 &  180.5  & - \\
H\_0A050\_F0\_128 & 2$\cdot$10$^{-4}$ & -    & -    & -    & 0        & 0.50  & 0        & -            & -   &  176.5  & - \\  \hline \\[-2.0ex]
M\_0A020\_W0.10\_128 & 2$\cdot$10$^{-4}$ & -   & 2$\cdot$10$^{-4}$ & 10$^{-6}$ & 0        & 0.20  & 1        & 0.02         & 0.1 &  36.1 & No \\
M\_0A020\_W0.15\_128 & 2$\cdot$10$^{-4}$ & -   & 2$\cdot$10$^{-4}$ & 10$^{-6}$ & 0        & 0.20  & 1        & 0.02         & 0.15 &  56.6  & Yes *\\
M\_0A020\_W0.25\_128 & 2$\cdot$10$^{-4}$ & -   & 2$\cdot$10$^{-4}$ & 10$^{-6}$ & 0        & 0.20  & 1        & 0.02         & 0.25 &  104.9  & Yes *\\ \hline \\[-2.0ex]
M\_0A020\_F0\_128    & 2$\cdot$10$^{-4}$ & -   & 2$\cdot$10$^{-4}$ & 10$^{-6}$ & 0        & 0.20  & 0        & -         & - &  70.5  & No \\ 
M\_0A020\_F0.5\_128  & 2$\cdot$10$^{-4}$ & -   & 2$\cdot$10$^{-4}$ & 10$^{-6}$ & 0        & 0.20  & 0.5     & 0.02         & 0.2 &   73.0  & Yes *\\
M\_0A020\_F1.5\_128 & 2$\cdot$10$^{-4}$ & -   & 2$\cdot$10$^{-4}$ & 10$^{-6}$ & 0        & 0.20   & 1.5      & 0.02         & 0.2 &  85.0  & Yes *\\ \hline \\[-2.0ex]
256$^3$            & $\nu$    & $\chi$ & $\eta$   & B$_0$    & $\Omega$ & $A$ & $\phi_0$ & $\Delta$t & R & Re  & Dynamo   \\ \hline \\[-2.0ex]
M\_0A010      & 2$\cdot$10$^{-4}$ & -      & 2$\cdot$10$^{-4}$ & 10$^{-6}$ & 0        & 0.10  & 1        & 0.02         & 0.2 &  49.8  & Yes *\\
M\_0A015      & 2$\cdot$10$^{-4}$ & -      & 2$\cdot$10$^{-4}$ & 10$^{-6}$ & 0        & 0.15  & 1        & 0.02         & 0.2 &  63.5  & Yes *\\
H\_0A020      & 2$\cdot$10$^{-4}$ & -      & -    & -    & 0        & 0.20  & 1        & 0.02         & 0.2 &  78.9  & - \\
M\_0A020      & 2$\cdot$10$^{-4}$ & -      & 2$\cdot$10$^{-4}$ & 10$^{-6}$ & 0        & 0.20  & 1        & 0.02         & 0.2 &  78.9  & Yes *\\
MB\_0A020     & 2$\cdot$10$^{-4}$ & 2$\cdot$10$^{-4}$   & 2$\cdot$10$^{-4}$ & 10$^{-6}$ & 0        & 0.50  & 1     & 0.02         & 0.2 &  76.9  & Yes *\\
M\_0A050      & 2$\cdot$10$^{-4}$ & -      & 2$\cdot$10$^{-4}$ & 10$^{-6}$ & 0        & 0.50  & 1        & 0.02         & 0.2 &  181.1  & Yes *\\ \hline \\[-2.0ex]
512$^3$            & $\nu$    & $\chi$ & $\eta$   & B$_0$    & $\Omega$ & $A$ & $\phi_0$ & $\Delta$t & R  & Re  & Dynamo    \\ \hline \\[-2.0ex]
M\_0A020\_512   & 2$\cdot$10$^{-4}$ & -      & 2$\cdot$10$^{-4}$ & 10$^{-6}$ & 0        & 0.20  & 1        & 0.02       & 0.2  &   78.5  & Yes *\\ \hline \hline
\end{tabular}
\label{table sinusoidal shear}
\tablefoot{The last column indicates whether or not a dynamo effect is observed. Simulations with * in the last column are analyzed in Table \ref{Dynamo growths}.}
\end{table*}

\begin{table*}[ht]
\scriptsize
\centering
\caption{Analysis of selected dynamo runs from Table \ref{table sinusoidal shear}.}
\begin{tabular}{ccccccccccccc}
\hline \hline \\[-2.0ex]
                     & A    & $\phi$ & R    & r ($t_{turn}^{-1}$) & r$_{\omega}$ ($t_{turn}^{-1}$) & r$_{\omega,sat}$ ($t_{turn}^{-1}$) & t$_{turn}$      & k$_{\omega}$/k$_{f}$      & Re (Rm)  & Re$_{\omega}$   & u$_{rot}$/u$_{tot,init}$  & u$_{rot}$/u$_{tot}$ \\  \hline  \\[-2.0ex]
M\_0A020\_W0.15\_128 & 0.20 & 1      & 0.15 & 0.0286              & 0.00918                        & 0.00120     & 0.50 &  0.071  & 56.6 &  4.01   & 0.87  &  0.91                        \\
M\_0A020\_128        & 0.20 & 1      & 0.2  & 0.0296              & 0.00983                        & 0.00113   &  0.62 &  0.092  & 80.2  & 7.37   & 0.78  &  0.81     \\
M\_0A020\_W0.25\_128 & 0.20 & 1      & 0.25 & 0.0235              & 0.00853                        & 0.000323    & 0.74 &  0.112  & 104.9 &  11.76  & 0.73  &  0.76   \\ 
M\_0A020\_F0.5\_128  & 0.20 & 0.5    & 0.2  & 0.0290              & 0.00875                        & 0.000644    & 0.68 &  0.096  & 74.1 &  7.10  & 0.91  &  0.92                          \\
M\_0A020\_F1.5\_128  & 0.20 & 1.5    & 0.2  & 0.0239              & 0.00857                        & 0.000333    & 0.58 &  0.099  & 85.8 &  8.49   & 0.70  &  0.75                       \\ \hline \\[-2.0ex]

M\_0A010             & 0.10 & 1      & 0.2  & 0.0155              & 0.00756                        & 0.000126 &  1.00  &   0.073  &  49.8  &  3.61  &  0.51  &  0.68                              \\
M\_0A015             & 0.15 & 1      & 0.2  & 0.0220              & 0.00944                        & 0.000304  & 0.79  &   0.085  &  63.5  &  5.40  &   0.70  &  0.76                                        \\
M\_0A020             & 0.20 & 1      & 0.2  & 0.0262              & 0.00969                        & 0.000379 &  0.63  &   0.091  &  78.9  &  7.19  &  0.81  &  0.86                        \\
MB\_0A020            & 0.20 & 1      & 0.2  & 0.0316              & 0.00925                        & 0.000802 &  0.65  &   0.093  &  76.9  &  7.15  &  0.85  &  0.88                            \\ 
M\_0A050             & 0.50 & 1      & 0.2  & 0.0510              & 0.0143                         & 0.00135 &   0.28  &   0.103  &  181.1  &  18.74  &   0.96  &  0.97                                        \\ \hline \\[-2.0ex]
M\_0A020\_512        & 0.20 & 1      & 0.2  & 0.0264              & 0.0107                         & -   &   0.64 &   0.090 &   78.5 &   7.10  &  0.83  &   -                                
 \\ \hline \hline
\end{tabular}
\label{Dynamo growths}
\tablefoot{We indicate the growth rates for: E$_{mag}$ (r), $\omega_{rms}$ (r$_{\omega}$) and $\omega_{rms}$ during the the winding phase (r$_{\omega,sat}$); see Fig. \ref{Dynamo growth time series} for the time evolution. The other columns are similar to the ones found in Table \ref{table barotropic} taken as an average before the dynamo instability, with the exception of the last two, which includes the Helmholtz decomposition averages before and after the instability. These are all the dynamo runs in Table \ref{table sinusoidal shear}, except M\_0A010\_128, M\_0A015\_128, and M\_0A050\_128, because of similar parameters with lower resolution, and except M\_0A100\_128, because the system crashed after the flow became supersonic.}
\end{table*}

In Fig. \ref{Hydro vs dynamo instability}, we show the evolution of the average magnetic energy density, $E_{mag}=\langle B^2\rangle /2\mu_0$, the average turbulent kinetic energy density, $E'_{kin}=\langle \rho u'^2\rangle /2,$ (i.e., neglecting the shearing contributions) and the root mean square of the turbulent vorticity $\mathbf{\omega'}=\nabla\times \mathbf{u}'$, for one MHD and one HD representative runs. All averages are taken over the entire domain. We notice that $E'_{kin}$ and $\omega'_{rms}$ grow almost exactly the same way in the HD and MHD cases during the initial instability. Then the MHD run shows SSD shortly after the vorticity dynamo. When the magnetic field increases to a dynamical regime, the MHD run sees a decrease in $E'_{kin}$ and $\omega'_{rms}$, until we almost reach equipartition between the magnetic and the turbulent kinetic energies. If, instead, the shear velocity is included, it dominates the total kinetic energy by a factor of a few, the value depending on the shear amplitude.

Such an exponential amplification of vorticity, or vorticity dynamo, was initially predicted by \cite{BlackmanChou1997ApJ} who, however, showed how an helical forcing was needed. The effect we see in our simulations was predicted by \cite{Elperinetal2003} for non-helical homogeneous turbulence with a mean velocity shear. It was then observed in the HD case by \cite{Kayplaetal2009} in a setup with linear shear and plane wave forcing. Since the Kelvin-Helmholtz is a kinetic instability, the MHD case leads to SSD just after the vortical structures are formed: the magnetic fields are twisted and stretched by advection, until the magnetic energy is strong enough to provide feedback on the fluid, slightly reducing the mean kinetic energy density. At a later stage, also a large-scale component of the magnetic field is amplified: this can be interpreted as the shear-current effect described by \cite{Igor2003PhRvE..68c6301R}, where an electromotive force proportional to vorticity is produced.

\begin{figure}[t]
\centering
\includegraphics[width=\hsize]{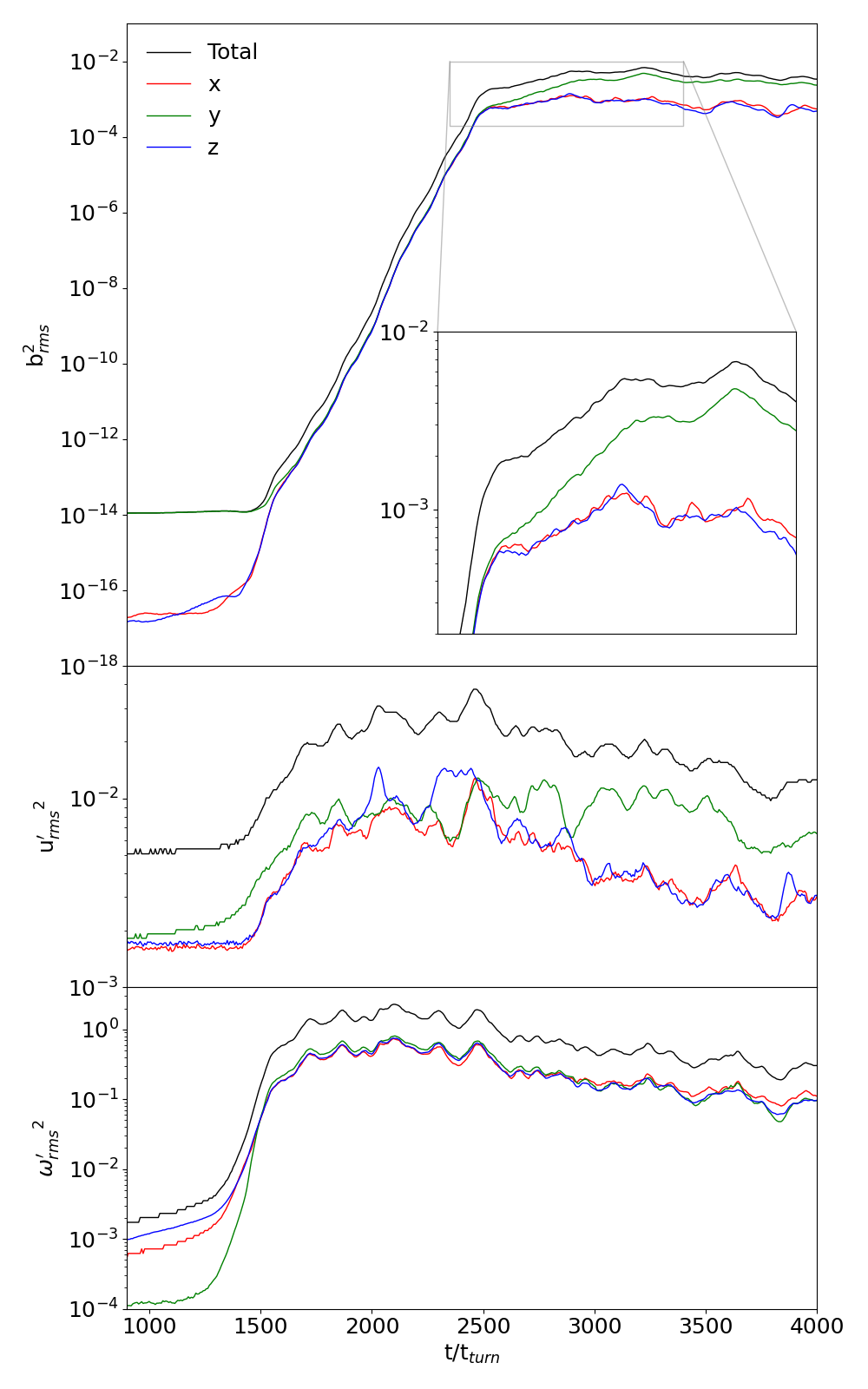}
\caption{Vector component evolution for the magnetic field, turbulent velocity and turbulent vorticity of the M\_0A020 run. The $y$ direction of the magnetic field shows an evolution different from the other components before and after, but not during, the dynamo phase. This is only slightly observed in the turbulent velocity field, while vorticity shows no preferential direction at all.}
\label{Winding}
\end{figure}

\begin{figure}[h]
\centering
\includegraphics[width=\hsize]{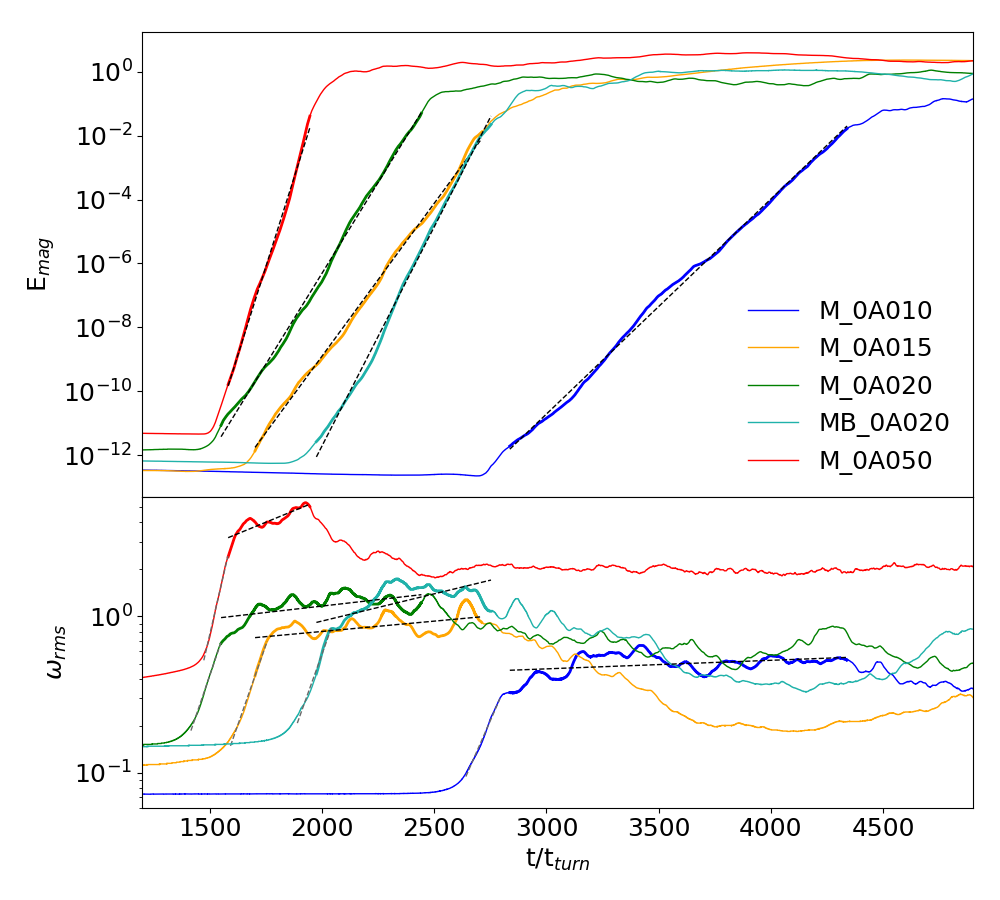}
\caption{Evolution of the volume-integrated magnetic energy and the  root mean square of the vorticity, $\omega_{rms}$, for some dynamo runs (as in the legend). The dynamo and vorticity growth rates increases with the shearing amplitude. The baroclinicity slightly enhances the dynamo growth rate even though it takes longer for the instability to start. The vorticity keeps growing after its exponential growth, during the magnetic dynamo phase. It then decays to lower values. The straight lines indicate the fits we obtained for the growth rates of the dynamo, r, of the vorticity dynamo, r$_{\omega}$, and of the growth of the vorticity during the first part of its saturation phase, r$_{\omega, sat}$. Their values are reported in Table \ref{Dynamo growths}.}
\label{Dynamo growth time series}
\end{figure}

\begin{figure*}[h]
\centering
\includegraphics[width=0.98 \hsize]{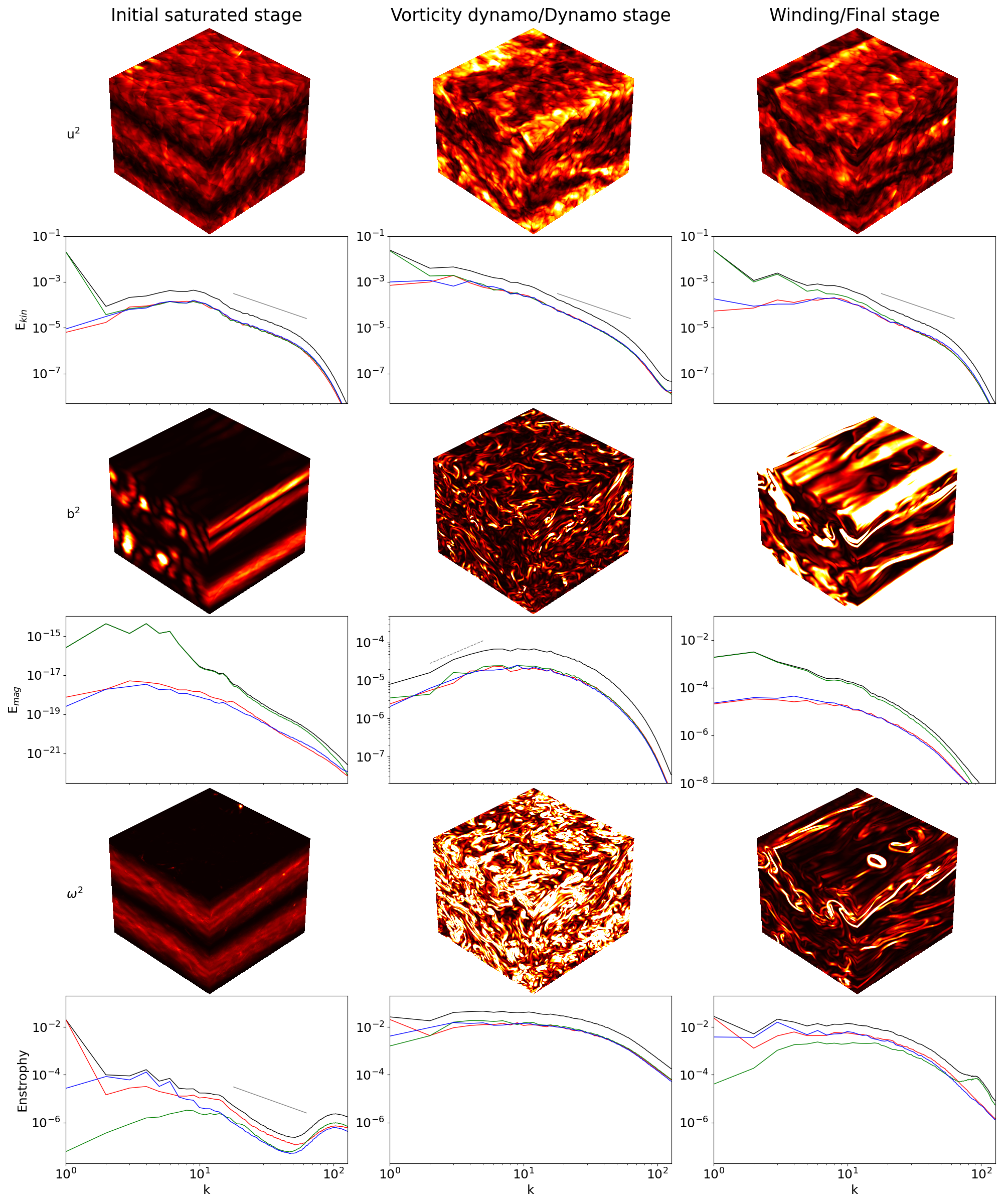}
\caption{Spectra and snapshots over three faces of the domain, for the squares of velocity, $u^2$, magnetic field, $b^2$, and vorticity, $\omega^2$, of the run M\_0A020. We display three different phases of the evolution of the system, evolving over time from left to right. Color bars have different ranges to allow a better visualization: for $u^2$ the range is (0,0.2); for $b^2$ the range is (0,1e-13) for the first snapshot and (0,0.015) for the others; for $\omega^2$ the range is (0,0.2) for the first snapshot and (0, 2.0) for the others, all in code units. The $x$, $y$ and $z$ components of the spectra correspond to the red, green and blue lines, respectively. Slope lines in the kinetic energy spectra are proportional to $k^{-2}$ before the dynamo and get a bit steeper afterwards. Magnetic energy spectra take a typical Kazantsev $k^{3/2}$ slope during the kinematic phase of the dynamo \citep{Kazantsev1968}. We note that the scale of the kinetic energy and vorticity spectra (or enstrophy spectra) are kept constant, but the magnetic ones are not, in order to follow the magnetic field growth.}
    \label{Spectra and snapshots sinusoidal shear}
\end{figure*}

In the presence of a differential velocity profile and once a sufficiently large magnetic field exists, we observe a relevant effect known as winding (e.g., \citealt{fujisawa15}). Given such a shear profile as ours, $u_y^S(z)$, the advective term of the induction equation provides an increase of the $B_y$ component:

\begin{equation}
\frac{\partial \mathbf{B}}{\partial t} = \nabla \times (\mathbf{u} \times \mathbf{B}) \quad \rightarrow \quad \frac{\partial B_y}{\partial t} = - B_z \frac{\partial u^S_y}{\partial z}~.
\end{equation}
This winding behavior leads to a linear growth of the magnetic field in the shearing direction which will stop when magnetic resistivity and/or dragging become important; namely: when the Lorentz force becomes relevant and the system non-linear. This can be seen for the model M\_0A020 in the upper plot of Fig. \ref{Winding}, where during the time of dynamo growth each component of the magnetic field is equally important, until the non-linear problem sets in. Although our setup is more complex, as the potential forcing produces flow in all directions, this results in different growth of the field in the y direction during the latest stage of the field growth. In the middle and bottom panels of Fig. \ref{Winding}, we also show the root mean square of turbulent vorticity and turbulent velocity. They show a mild preferred contribution only in the $y$-component of the turbulent velocity (probably for the feedback of the strong magnetic field).

In Table \ref{Dynamo growths}, we show the different growth rates (in units $t_{turn}^{-1}$) obtained for all dynamo runs. In Fig. \ref{Dynamo growth time series} we also show the volume-integrated magnetic energy $E_{mag}$ and the root mean square of the vorticity $\omega_{rms}$ evolution for some of the 256$^3$ runs. We have found that the growth rate of the magnetic field is roughly proportional to the shear rate, similar to what was found by \cite{Kayplaetal2009sinusoidal}. The dynamo growth seems to have saturated already in the 256$^3$ run, but the vorticity growth seems to vary more. The 512$^3$ run was stopped after the dynamo growth for computational reasons, thus we do not show the last slope. Baroclinicity does not change much the vorticity instability growth, but allows for a more rapid dynamo growth. For the runs with constant amplitude ($A$=0.2) and increasing resolution there is little difference between the 256$^3$ run and the 512$^3$, which indicates numerical convergence. Comparing the 128$^3$ dynamo runs with different forcing, we may notice that the dynamo and vorticity growth rates increase with both $\phi_0$ and $R$, but when they exceed a certain value, both rates decrease. This is probably caused by the flow reaching a transonic regime and decreasing the growth efficiency \citep{Haugen&Barandenburg2004, Federrathetal2011, Schleicheretal2013}.

We did not reach dynamo for the cases with a shear amplitude $A\lesssim 0.1$, which can be translated as a critical Reynolds value of  Re$_{crit}\sim 50$, and a minimal rotational component of the flow of about 50\% which in our models corresponds to $k_{\omega}/k_{f,crit}$ of 0.073. Although these specific values might depend on the resolution, this transition from non-dynamo to dynamo generating flows was not observed in the rigidly rotating cases.

Finally, in Fig. \ref{Spectra and snapshots sinusoidal shear} we represent the spectra and snapshots of M\_0A020 to give a more visual and detailed description of the instability. This figure summarizes the evolution of the main quantities during the three main evolutionary stages of our simulations.

\section{Possible astrophysical applications} \label{Astrophysical applications}
The presented model is rather general and can therefore be applied to several astrophysical environments. In particular, the implemented forcing can be interpreted as the expansion waves coming from SNe, the main forcing in ISM. Although SNe tend to be highly supersonic and spherically asymmetric, the expansion waves under consideration here can be thought as the long-time expanding wave of the SN remnant.

We can interpret our box as a small cube with side 500 pc inside a galaxy (i.e., a resolution of about 2 pc). The forcing width of 1/10 of the box corresponds to 50 pc, slightly larger but of the same order of magnitude of some SN remnants. We can take the reference value for density of 10$^{-23}$ g/cm$^{3}$, and the speed of sound of 10 km/s. Galactic rotation curves lead to a shearing amplitude of about $5$ km/s for a 500 pc radial distance, thus, it is on the same order of magnitude, $A=0.2,$ for the sinusoidal shear in our model.

The estimated SN rate of 2-3 SN per century per galaxy \citep{murphey21} can be then translated to a 3-4 SN/Myr per (500 pc)$^3$-box, assuming a galactic volume of the order of 10$^{12}$ pc$^3$. The value $\Delta t$ of 0.02 leads to a rate of $\sim 6$ SN/My per (500 pc)$^3$-box. As an example, for the M\_0A020 run, these units lead to a $u_{rms}$ of the order of 2 km/s and a mean magnetic field of 10 ${\mu}$G, which is of the order of the estimated galactic one \citep[e.g.][]{jansson12,BN2023ARAA}.

Therefore, the results can be indeed of direct application for ISM. More realistic simulations could consider density variations and non spherically symmetric expansion waves (which might lead easier to SSD). Finally, the shear here considered could be refined in order to replicate for instance the rotational curve of the galaxy or the shear corresponding to the spiral arms.

\section{Conclusions}
\label{sec:conclusions}
In this work, we investigate the possibility of having vorticity production and dynamo action driven by a purely curl-free forcing of the velocity field. We ran numerical models that employed an irrotational forcing, which does not lead to any growth of the vorticity  neither in the HD nor in the MHD scenarios, if it is acting independently. However, a vortical flow is produced when the forcing interacts with a solid body rotation, in the presence of baroclinicity, or of a background shear. In the case of a rotating and baroclinic system we have not found any dynamo action within the explored parameter space, neither with an initial random seed magnetic field, nor with an initial uniform field. This result, therefore, seems not to be dependent on the topology of the initial field.

Instead, in the presence of a background, sinusoidal shearing velocity, we could observe an amplification of the initial seed magnetic field as a consequence of a dynamo process, both in the barotropic and in the baroclinic cases. The main novelty in the presented work is to show how such a dynamo phase occurs after the onset of a hydrodynamical instability driving an exponential growth of the vorticity. After the vorticity has grown, the magnetic energy is amplified too, approaching equipartition with the turbulent kinetic energy. This exponential amplification of vorticity occurs both in the purely HD case and in the MHD one, starting from small scales and then spreading up to the scale of the box. Similarly, the magnetic field is amplified first on small scales, but the inverse cascade makes the large scales dominate during the saturation phase of the simulations. Such large-scale field is related to the winding process and made visible after the equipartition. This winding enhances the field in the shearing direction, but the turbulent vorticity and velocity remain isotropic.

If the baroclinic term is at work, we noticed how the vorticity growth takes place at a slightly later time, with a growth rate similar to the barotropic case. Instead, the magnetic field grows at a bit faster rate.

Our results cannot firmly exclude that a dynamo is produced with irrotational forcing, rotation, and no shear, as, for example, seen by \citealt{Seta2022MNRAS.514..957S} using a different Fourier-based forcing in a multi-phase MHD setup with the Flash code. We can, however, affirm that in our setup, such a putative dynamo could only be produced in a regime of higher Re than what tested in this work (up to a few hundreds). In any case, the critical threshold would therefore be much larger than what seen in the shear case (Re$_{crit}\sim 50$).

Possible follow-up works in the context of dynamo triggered by an irrotational forcing could be focused on the exploration of higher Re and resolutions, as well as the inclusion of a stratified shearing medium. Such results are ultimately applicable to study galactic environments and other astrophysical scenarios, such as accretion disks and planetary or stellar atmospheres.

\begin{acknowledgements}
This work has been carried out within the framework of the doctoral program in Physics of the Universitat Autònoma de Barcelona and it is partially supported by the program Unidad de Excelencia María de Maeztu CEX2020-001058-M. DV and AE are supported by the European Research Council (ERC) under the European Union’s Horizon 2020 research and innovation programme (ERC Starting Grant "IMAGINE" No. 948582, PI: DV). FDS acknowledges support from a Marie Curie Action of the European Union (Grant agreement 101030103). The authors acknowledge support from ``Mar\'ia de Maeztu'' award to the Institut de Ciències de l'Espai (CEX2020-001058-M). We are grateful to Axel Brandenburg, Maarit Korpi-Lagg, Frederick Gent, Matthias Rheinhardt, Eva Ntormousi, Amit Seta and Claudia Soriano-Guerrero for fruitful discussions. We also want to acknowledge the whole Pencil Code community for support. AE and FDS gratefully acknowledge NORDITA for hospitality during the Program “Magnetic field evolution in low density or strongly stratified plasmas” in May 2022, where part of this work was performed.
\end{acknowledgements}

\bibliographystyle{aa}
\bibliography{biblio}

\begin{appendix}
\section{Spectra} \label{Spectra}

Spectra of a continuous variable ($\tilde{u}$) can be calculated by numerically obtaining the 3D Fourier transform:
\begin{equation}
\label{Fourier transform}
\begin{aligned}
  \tilde{\Vec{u}}(\Vec{k}) = \frac{1}{(2 \pi)^3} \int_0^L\int_0^L\int_0^L \Vec{u} (\Vec{r}) e^{- i \Vec{k} \cdot \Vec{r}} dx~dy~dz \approx \\ \approx \frac{1}{N^3} \sum_{p=0}^{N-1}\sum_{q=0}^{N-1}\sum_{r=0}^{N-1} u(x_p, y_q, z_r)e^{-i k_x x_p} e^{-i k_y y_q} e^{-i k_z z_r}
\end{aligned}
,\end{equation}
where $N$ is the numbers of points along each box direction, $L=2\pi$ is the length of each box side, $(x_p,y_q,z_r)$ indicate the discrete points of the mesh, and $k_i$ are the set of integer wavenumbers related to each directions, each defined in the range $[1,N/2]$. The three dimensional (3D) spectrum is defined as:
\begin{equation}
\label{3D Spectra}
  P(k) = \frac{1}{2} \tilde{\Vec{u}}(\Vec{k}) \tilde{\Vec{u}}^*(\Vec{k}) \quad \quad \text{where} \quad \quad k = \sqrt{k_x^2 + k_y^2 + k_z^2}
,\end{equation}
where the three-dimensional dependence of the spectrum on $\vec{k}$ is reduced to a function of the modulus $k$ alone, by rebinning in the Fourier space.

This has been used for the velocity (kinetic energy spectra, neglecting the small density deviations from unity), magnetic field (magnetic energy spectra), and vorticity (enstrophy spectra).

\section{Helmholtz decomposition} \label{Helmholtz decomposition}

Helmholtz showed that any vector field $\mathbf{F}$ which vanishes suitably quickly at infinity can be decomposed into  irrotational (longitudinal, purely divergent) and solenoidal (transverse) components. Therefore it can be expressed as the gradient and curl of a scalar and gradient potentials, respectively:
\begin{equation}
    \mathbf{F}= \mathbf{F}_{pot} + \mathbf{F}_{rot} = -\nabla \phi+\nabla\times\mathbf{A}~.
\end{equation}
Aided by the Pencil code available tools, we have used this decomposition to evaluate how much of the flow is irrotational, with no associated vorticity, and how much is rotational. To numerically obtain such decomposition we made use of the 3D Fourier transform to obtain the flow components in Fourier space: $\tilde{\Vec{u}}_{pot}(\Vec{k}) = \tilde{\Vec{u}}(\Vec{k}) \cdot \Vec{k}$, $\tilde{\Vec{u}}_{rot}(\Vec{k}) = \tilde{\Vec{u}}(\Vec{k}) \times \Vec{k}$. If we inverse transform these functions, we can obtain the decomposition. We can check both the percentage of the rotational flow and the error of the numerical procedure by comparing the squared volume quantities with the relation:
\begin{equation}
    \Vec{u}(\Vec{r}) = \Vec{u}_{pot}(\Vec{r}) + \Vec{u}_{rot}(\Vec{r}) \quad \rightarrow{} \quad \langle |\Vec{u}|^2\rangle = \langle |\Vec{u}_{pot}|^2\rangle + \langle |\Vec{u}_{rot}|^2\rangle
.\end{equation}
In the analysis, we have chosen only one single snapshot (once saturation is reached) to perform the decomposition, for computational practical reasons. We verified that averaging among many snapshots does indeed lead to the percentage of rotational flow oscillating between values that vary by no more than 5 \%.

\section{Linear shear} \label{Linear shear}

We could opt for the use of a linear shear of the form $\mathbf{u}^S = (0, Sx, 0)$. With this choice, the $yz$-faces of the box cannot satisfy periodic boundary conditions, thus shearing-periodic boundary conditions are needed: the $x$ direction is periodic with respect to positions in $y$ that shift in time:
\begin{equation*}
    f(-\frac{1}{2} L_x, y, z, t) = f(\frac{1}{2}L_x, y + L_x S t, z, t),
\end{equation*}
where $f$ is each of the evolving fields. This condition is routinely used in numerical studies of shear flows in Cartesian geometry since it was introduced by \cite{Wisdom&Tremaine1988} and \cite{Howleyetal1995}. These boundary conditions are known to potentially produce spurious vorticity at the boundary, which, also due to the irrotational nature of our forcing, we clearly notice, as we discuss below.

We show the tested cases in Table \ref{table shear} (see Table~\ref{table shears} for the diagnostics of some cases). For reasons of numerical stability, we set higher values of viscosity for the 128$^3$ tests with the linear shear term. This leads us to a lower Re compared to the most of the other shear-less simulations. Regarding the magnetic field evolution (last column of Table \ref{table shear}), in most cases we find a decay of magnetic energy, which depends on the shear parameter $S$. Only in cases with a low magnetic diffusivity ($2 \cdot 10^{-5}$), we observe some magnetic field exponential growth. However, we cannot draw any reliable conclusion about dynamo action, for the following reasons.

Looking at the trend between the vorticity indicator (at saturation) and the shear amplitude, $S$, $k_{\omega}/k_f$ = (0.108$\pm$0.001) + (0.151$\pm$0.005) $S$, we notice an nonphysical non-zero (and non-negligible) vorticity for vanishing shear, $S$, and we checked that it is not cured by increasing the resolution. This was already observed by \cite{DelSordoBrandenburg2011} and attributed to numerical artifacts due to the interaction between the expansion waves laying on both sides of the shearing boundary.

A visual inspection of the results reveals indeed that almost the entire vorticity and magnetic field production were near the shearing boundaries. As an attempt to mitigate this artifact, we spatially constrained the expansion waves in the middle half of the box (i.e., as far away as possible from the shearing boundaries). This indeed resulted in a more spatially uniform and slower growth of vorticity and magnetic field, but we still could not keep the system from being affected by the aforementioned spurious contributions at the shearing boundaries after a certain time. Higher values of $S$ or the addition of rotation seem to amplify some magnetic field in the middle of the box, but the boundary contribution was still dominant.
This numerical issue has led us to use the more complex sinusoidal shearing profile which allows simple periodic boundary conditions, with no spurious effects.

\begin{table}[t]
\centering
\scriptsize
\caption{MHD simulations with linear shear and the barotropic EoS.}
\begin{tabular}{@{}cccccc@{}}
\hline  \hline \\[-2.0ex]
128$^3$            & $\nu$   & $\eta$     & $\Omega$ & $S$ & B$_{growth}$   \\ \hline \\[-2.0ex]
M\_0S00\_128 & 2$\cdot$10$^{-3}$  & 2$\cdot$10$^{-3}$ &  0        & 0    & No \\
M\_0S01\_128 & 2$\cdot$10$^{-3}$  & 2$\cdot$10$^{-3}$ & 0        & 0.01  &  No  \\
M\_0S05\_128 & 2$\cdot$10$^{-3}$  & 2$\cdot$10$^{-3}$ &  0        & 0.05   & No  \\
M\_0S10\_128 & 2$\cdot$10$^{-3}$  & 2$\cdot$10$^{-3}$ &  0        & 0.1    & No  \\
M\_0S15\_128 & 2$\cdot$10$^{-3}$  & 2$\cdot$10$^{-3}$ &  0        & 0.15  & No  \\
M\_0S20\_128 & 2$\cdot$10$^{-3}$  & 2$\cdot$10$^{-3}$ &  0        & 0.2    & No  \\
M\_0S30\_128 & 2$\cdot$10$^{-3}$  & 2$\cdot$10$^{-3}$ &  0        & 0.3    & No  \\
M\_0S40\_128 & 2$\cdot$10$^{-3}$  & 2$\cdot$10$^{-3}$ &  0        & 0.4    & No  \\
M\_0S50\_128 & 2$\cdot$10$^{-3}$  & 2$\cdot$10$^{-3}$ &  0        & 0.5    & No  \\  \hline \\[-2.0ex]
M\_0S00\_Pm\_128 & 2$\cdot$10$^{-3}$ & 2$\cdot$10$^{-5}$ &  0        & 0      & No \\
M\_0S01\_Pm\_128 & 2$\cdot$10$^{-3}$ & 2$\cdot$10$^{-5}$ &  0        & 0.01   & Yes  \\
M\_0S05\_Pm\_128 & 2$\cdot$10$^{-3}$ & 2$\cdot$10$^{-5}$ &  0        & 0.05   & Yes  \\
M\_0S10\_Pm\_128 & 2$\cdot$10$^{-3}$ & 2$\cdot$10$^{-5}$ &  0        & 0.1    & Yes  \\
M\_0S15\_Pm\_128 & 2$\cdot$10$^{-3}$ & 2$\cdot$10$^{-5}$ &  0        & 0.15   & Yes  \\
M\_0S20\_Pm\_128 & 2$\cdot$10$^{-3}$ & 2$\cdot$10$^{-5}$ &  0        & 0.2    & Yes  \\
M\_0S30\_Pm\_128 & 2$\cdot$10$^{-3}$ & 2$\cdot$10$^{-5}$ &  0        & 0.3    & Yes  \\
M\_0S40\_Pm\_128 & 2$\cdot$10$^{-3}$ & 2$\cdot$10$^{-5}$ &  0        & 0.4    & Yes  \\
M\_0S50\_Pm\_128 & 2$\cdot$10$^{-3}$ & 2$\cdot$10$^{-5}$ &  0        & 0.5    & Yes  \\ \hline \\[-2.0ex]
M\_.2S10\_Pm\_128 & 2$\cdot$10$^{-3}$ & 2$\cdot$10$^{-5}$ &  0.2      & 0.1    & Yes  \\
M\_2S10\_Pm\_128 & 2$\cdot$10$^{-3}$ & 2$\cdot$10$^{-5}$ &  2        & 0.1   & Yes  \\
M\_.2S20\_Pm\_128 & 2$\cdot$10$^{-3}$ & 2$\cdot$10$^{-5}$ &  0.2      & 0.2   & Yes  \\
M\_2S20\_Pm\_128 & 2$\cdot$10$^{-3}$ & 2$\cdot$10$^{-5}$ &  2        & 0.2  & Yes  \\
M\_.2S30\_Pm\_128 & 2$\cdot$10$^{-3}$ & 2$\cdot$10$^{-5}$ &  0.2      & 0.3    & Yes  \\
M\_2S30\_Pm\_128 & 2$\cdot$10$^{-3}$ & 2$\cdot$10$^{-5}$ &  2        & 0.3    & Yes  \\ \hline \\[-2.0ex] \hline 
\end{tabular}
\tablefoot{In all the run, $B_0=10^{-6}$, $\phi_0=1$, $\Delta$t$=0.02$, $R=0.2$. The magnetic field growth (last column) seen in some of them is dominated by a spurious boundary effect.}
\label{table shear}
\end{table}

\begin{table}
\centering
\scriptsize
\caption{Diagnostic magnitudes for runs from Table \ref{table shear} which include linear shear, no rotation, and with $Pm=100$.}
\begin{tabular}{cccccc}
\hline \hline \\[-2.0ex]
       & $S$ & t$_{turn}$      & k$_{\omega}$/k$_{f}$      & Re   & Re$_{\omega}$   \\ \hline  \\[-2.0ex]         
M\_0S00\_Pm\_128 & 0.00 & 1.99 & 0.00615 & 2.52 & 0.02 \\
M\_0S01\_Pm\_128 & 0.01 & 1.99 & 0.11165 & 2.52 & 0.28 \\
M\_0S05\_Pm\_128 & 0.05 & 2.00 & 0.11509 & 2.49 & 0.29 \\
M\_0S10\_Pm\_128 & 0.10 & 2.03 & 0.12138 & 2.46 & 0.3  \\
M\_0S15\_Pm\_128 & 0.15 & 2.06 & 0.12949 & 2.43 & 0.32 \\
M\_0S20\_Pm\_128 & 0.20 & 2.11 & 0.13702 & 2.37 & 0.32 \\
M\_0S30\_Pm\_128 & 0.30 & 2.17 & 0.15462 & 2.32 & 0.34 \\
M\_0S40\_Pm\_128 & 0.40 & 2.29 & 0.16855 & 2.19 & 0.37 \\
M\_0S50\_Pm\_128 & 0.50 & 2.34 & 0.18544 & 2.14 & 0.39 \\ \hline  \hline 
\end{tabular}
\label{table shears}
\end{table}

\end{appendix}

\end{document}